\documentclass[sigconf]{acmart}

\usepackage{graphicx}
\usepackage{amsmath}
\usepackage{booktabs}
\usepackage{siunitx}
\usepackage{url}

\usepackage{caption}
\usepackage{subcaption}

\renewcommand\footnotetextcopyrightpermission[1]{} 
\begin{document}

\title{Corrected Evaluation Results of the NTCIR WWW-2, WWW-3, and WWW-4 English Subtasks}


%


\author{Tetsuya Sakai, Sijie Tao}
\affiliation{Waseda University, Japan}
\email{tetsuyasakai@acm.org}
\email{tsjmailbox@ruri.waseda.jp}

\author{Maria Maistro}
\affiliation{University of Copenhagen, Denmark}
\email{mm@di.ku.dk}

\author{Zhumin Chu}
\affiliation{Tsinghua University, P.R.C.}
\email{chuzm19@mails.tsinghua.edu.cn}

\author{Yujing Li, Nuo Chen}
\affiliation{Waseda University, Japan}
\email{liyujing@ruri.waseda.jp}
\email{pleviumtan@toki.waseda.jp}

\author{Nicola Ferro}
\affiliation{University of Padua, Italy}
\email{ferro@dei.unipd.it}

\author{Junjie Wang}
\affiliation{Waseda University, Japan}
\email{wjj1020181822@toki.waseda.jp}

\author{Ian Soboroff}
\affiliation{NIST, USA}
\email{ian.soboroff@nist.gov}

\author{Yiqun Liu}
\affiliation{Tsinghua University, P.R.C.}
\email{yiqunliu@tsinghua.edu.cn}

\begin{abstract}
Unfortunately, the official English (sub)task results 
reported in the NTCIR-14 WWW-2,
NTCIR-15 WWW-3, and
NTCIR-16 WWW-4 overview papers
are incorrect due to noise in the official qrels files; this paper reports results
based on the corrected qrels files.
The noise is due to a fatal bug in the backend of our relevance assessment interface.
More specifically,
at WWW-2, WWW-3, and WWW-4,
two versions of pool files were created for each English topic:
a PRI (``prioritised'') file, which uses the NTCIRPOOL script to prioritise likely relevant documents,
and a RND (``randomised'') file, which randomises the pooled documents.
This was done for the purpose of studying the effect of document ordering for relevance assessors.
However, the programmer who wrote the interface backend
assumed that a combination of 
a topic ID and a document rank in the pool file
uniquely determines a document ID;
this is obviously incorrect as we have two versions of pool files.
The outcome is that
all the PRI-based relevance labels for the WWW-2 test collection are incorrect
(while all the RND-based relevance labels are correct),
and all the RND-based relevance labels for the WWW-3 and WWW-4 test collections are incorrect
(while all the PRI-based relevance labels are correct).
This bug was finally discovered at the NTCIR-16 WWW-4 task when the first seven authors of this paper served as Gold assessors
(i.e., topic creators who define what is relevant)
and closely examined the disagreements with Bronze assessors (i.e., non-topic-creators; non-experts).
We would like to apologise to the WWW participants and the NTCIR chairs for the inconvenience and confusion caused due to this bug.
\end{abstract}

\maketitle
\pagestyle{plain} 


\section{Introduction}

This paper corrects the official English (sub)task results 
of the NTCIR-14 WWW-2,
NTCIR-15 WWW-3, and
NTCIR-16 WWW-4 tasks~\cite{mao19,sakai20www3,sakai22www4}.
The corrected qrels files and the evaluation results are available from
\url{http://sakailab.com/www234corrected}.
We would like to apologise to the WWW participants and the NTCIR chairs for the inconvenience and confusion caused due to this bug.
More specifically,
the English run results of the following participants
are incorrect due to the bug:
WWW-2 participants (MPII~\cite{yates19}, RUCIR~\cite{yang19}, SLWWW~\cite{xiao19}, and THUIR~\cite{zheng19}),
WWW-3 participants (ESTUCeng~\cite{aydin20}, KASYS~\cite{shinden20}, NAUIR~\cite{liang20}, RUCIR~\cite{zuo20}, SLWWW~\cite{muraoka20}, THUIR~\cite{chu20}, Technion~\cite{raiber20}, and mpii~\cite{li20}), and
WWW-4 participants (KASYS~\cite{usuha22}, SLWWW~\cite{ubukata22}, and THUIR~\cite{yang22}).
Apologies again.

\section{NTCIR-14 WWW-2 English Subtask Results}\label{s:www2}

This section corrects the English subtask results reported in the WWW-2 overview paper~\cite{mao19}.

The WWW-2 English topic set contains 80 topics,
and each topic was handled by two assessors.
For 27 topics\footnote{
0001, 0004, 0007, 0010, 0013, 0016, 0019, 0022, 0025, 0028, 0031, 0034, 0037, 0040, 0043, 0046, 0049, 0052, 0055, 0058, 0061, 0064, 0067, 
0070, 0073, 0076, 0079},
both assessors used a PRI file (``PRI-PRI topics'');
for 26 topics\footnote{
0003, 0006, 0009, 0012, 0015, 0018, 0021, 0024, 0027, 0030, 0033, 0036, 0039, 0042, 0045, 0048, 0051, 0054, 0057, 0060, 0063, 0066, 0069, 
0072, 0075, 0078},
both assessors used a RND file (``RND-RND topics'');
for the remaining 27 topics\footnote{
0002, 0005, 0008, 0011, 0014, 0017, 0020, 0023, 0026, 0029, 0032, 0035, 0038, 0041, 0044, 0047, 0050, 0053, 0056, 0059, 0062, 0065, 0068, 
0071, 0074, 0077, 0080},
one assessor used a PRI file while the other assessor used a RND file (``PRI-RND topics'').
At the WWW-2 task,
the official qrels file was formed by 
summing the two 3-point scale relevance labels for each topic;
thus a 5-point scale qrels file containing relevance levels of L0-L4 
was obtained. 
However, for this particular round of the WWW task,
the PRI-based relevance labels were incorrect
due to the aforementioned bug.
Hence, the combined relevance labels for the 27 PRI-PRI topics and the 27 PRI-RND topics are noisy.

After correcting the PRI-based relevance labels,
we formed a correct 5-point scale qrels file using the above method.
Table~\ref{t:www2assessments} shows 
the statistics of the corrected WWW-2 qrels file.

\begin{table}[t]
\begin{center}
 
\caption{NTCIR-14 WWW-2 English relevance assessment statistics. This table replaces
the ``English'' column in Table~8 of \citet{mao19}.}\label{t:www2assessments}
\begin{tabular}{c|r}
\hline
\#topics			&80\\
\#assessors/topic	&2\\
Pool depth		&50\\
\hline
L4-relevant		&907\\
L3-relevant		&2,437\\
L2-relevant		&4,462\\
L1-relevant		&6,358\\
L0				&13,463\\
\hline
\#docs pooled		&27,627\\
\hline
\end{tabular}

\end{center}
\end{table}

Tables~\ref{t:www2corrected-nDCG-Q}-\ref{t:www2corrected-nERR-iRBU}
show the WWW-2 English run rankings based on the corrected qrels file.
Tables~\ref{t:www2corrected-significance-nDCG-Q} and~\ref{t:www2corrected-significance-nERR-iRBU} show the corresponding statistical significance test results.
We can observe that:
\begin{itemize}
\item In terms of nDCG and Q, 
THUIR-E-CO-MAN-Base-3 
is statistically the best run in that it
statistically significantly outperforms eight/seven other runs
(Table~\ref{t:www2corrected-significance-nDCG-Q}).
\item In terms of nERR,
THUIR-E-CO-MAN-Base-\{3, 2\} and RUCIR-E-CO-PU-Base-2 
are statistically the best runs in that they all
statistically significantly outperform five other runs
(Tables~\ref{t:www2corrected-significance-nERR-iRBU}(a)).
\item In terms of iRBU (which was not considered at the WWW-2 task but discussed here for the sake of consistency with WWW-3 and WWW-4),
the top ten runs in Table~\ref{t:www2corrected-nERR-iRBU}(b) are statistically tied:
THUIR-E-CO-MAN-Base-3 through RUCIR-E-DE-PU-Base-1
all statistically significantly outperform four other runs
(Tables~\ref{t:www2corrected-significance-nERR-iRBU}(b)).
\end{itemize}

In the WWW-2 overview paper where the noisy qrels file was used~\cite{mao19},
THUIR-E-CO-MAN-Base-3 was statistically the best run in terms of nDCG,
while 
THUIR-E-CO-MAN-Base-2 was statistically the best run in terms of Q;
no statistically significant differences were observed in terms of nERR.
The original conclusion based on the above results 
was that ``\textit{runs from THUIR are the most effective.}''
Our corrected results do not contradict the above conclusion,
but more clearly points to THUIR-E-CO-MAN-Base-3
as the best English run of the WWW-2 task.

Table~\ref{t:www2corrected-similarities}
shows the system ranking similarity for each pair of evaluation measure in terms of Kendall's $\tau$,
using the corrected WWW-2 qrels file for evaluating the 20 WWW-2 runs.
It can be observed that the four system rankings are quite similar,
with iRBU (introduced at WWW-3, not at WWW-2) behaving slightly differently compared to the other three.

\begin{table*}[t]
\begin{center}
 
\caption{
Corrected results for the WWW-2 runs ($n=80$ topics).
This table replaces
the Mean nDCG and Mean Q scores in
Table~12 of \citet{mao19}.}\label{t:www2corrected-nDCG-Q}
\begin{tabular}{l|r||l|r}
\hline
Run							&Mean nDCG@10	&Run				&Mean Q@10\\
\hline
THUIR-E-CO-MAN-Base-3 	&0.4804 &THUIR-E-CO-MAN-Base-3 	&0.4681\\
THUIR-E-CO-MAN-Base-2 	&0.4608 &THUIR-E-CO-MAN-Base-2 	&0.4524\\
THUIR-E-CO-MAN-Base-1 	&0.4459 &THUIR-E-CO-MAN-Base-1 	&0.4358\\
RUCIR-E-CO-PU-Base-2 		&0.4402 &MPII-E-CO-NU-Base-2 		&0.4319\\
RUCIR-E-DE-PU-Base-4 		&0.4342 &RUCIR-E-CO-PU-Base-2 		&0.4310\\
MPII-E-CO-NU-Base-2 		&0.4329 &RUCIR-E-DE-PU-Base-4 		&0.4276\\
MPII-E-CO-NU-Base-1 		&0.4210 &MPII-E-CO-NU-Base-1 		&0.4157\\
MPII-E-CO-NU-Base-5 		&0.4093 &MPII-E-CO-NU-Base-5 		&0.4006\\
MPII-E-CO-NU-Base-3 		&0.4077 &MPII-E-CO-NU-Base-3 		&0.3969\\
baseline\_eng\_v1 				&0.4032 &baseline\_eng\_v1 				&0.3884\\
THUIR-E-CO-PU-Base-5 		&0.4032 &THUIR-E-CO-PU-Base-5 		&0.3884\\
MPII-E-CO-NU-Base-4 		&0.4022 &MPII-E-CO-NU-Base-4 		&0.3856\\
RUCIR-E-DE-PU-Base-3 		&0.3915 &THUIR-E-CO-PU-Base-4 		&0.3755\\
RUCIR-E-DE-PU-Base-1 		&0.3915 &RUCIR-E-DE-PU-Base-3 		&0.3734\\
THUIR-E-CO-PU-Base-4 		&0.3870 &RUCIR-E-DE-PU-Base-1 		&0.3734\\
RUCIR-E-DE-PU-Base-5 		&0.3336 &RUCIR-E-DE-PU-Base-5 		&0.3181\\
SLWWW-E-CO-NU-Base-1 	&0.3300 &SLWWW-E-CO-NU-Base-1 	&0.3153\\
ORG-MANUAL 				&0.2682 &ORG-MANUAL 				&0.2527\\
SLWWW-E-CO-NU-Base-4 	&0.2661 &SLWWW-E-CO-NU-Base-4 	&0.2400\\
SLWWW-E-CD-NU-Base-3 		&0.2661 &SLWWW-E-CD-NU-Base-3 	&0.2400\\
\hline
\end{tabular}

\end{center}
\end{table*}

\begin{table*}[t]
\begin{center}
 
\caption{
Corrected results for the WWW-2 runs ($n=80$ topics).
This table replaces
the Mean nERR scores in
Table~12 of \citet{mao19};
it also shows the Mean iRBU scores for the sake of consistency 
with our WWW-3 and WWW-4 results.
}\label{t:www2corrected-nERR-iRBU}
\begin{tabular}{l|r||l|r}
\hline
Run							&Mean nERR@10	&Run				&Mean iRBU@10\\
\hline
THUIR-E-CO-MAN-Base-3 	&0.6430 &THUIR-E-CO-MAN-Base-3 	&0.8698\\
THUIR-E-CO-MAN-Base-2 	&0.6422 &THUIR-E-CO-MAN-Base-2 	&0.8547\\
RUCIR-E-CO-PU-Base-2 		&0.6356 &RUCIR-E-CO-PU-Base-2 		&0.8530\\
RUCIR-E-DE-PU-Base-4 		&0.6021 &THUIR-E-CO-MAN-Base-1 	&0.8441\\
THUIR-E-CO-MAN-Base-1 	&0.6007 &MPII-E-CO-NU-Base-2 		&0.8355\\
MPII-E-CO-NU-Base-2 		&0.5842 &MPII-E-CO-NU-Base-5 		&0.8354\\
MPII-E-CO-NU-Base-4 		&0.5730 &RUCIR-E-DE-PU-Base-4 		&0.8348\\
MPII-E-CO-NU-Base-1 		&0.5705 &MPII-E-CO-NU-Base-3 		&0.8306\\
baseline\_eng\_v1 				&0.5582 &RUCIR-E-DE-PU-Base-3 		&0.8272\\
THUIR-E-CO-PU-Base-5 		&0.5582 &RUCIR-E-DE-PU-Base-1 		&0.8272\\
MPII-E-CO-NU-Base-5 		&0.5460 &MPII-E-CO-NU-Base-1 		&0.8123\\
MPII-E-CO-NU-Base-3 		&0.5316 &baseline\_eng\_v1 				&0.8099\\
RUCIR-E-DE-PU-Base-3 		&0.5288 &THUIR-E-CO-PU-Base-5 		&0.8099\\
RUCIR-E-DE-PU-Base-1		&0.5288 &MPII-E-CO-NU-Base-4 		&0.7988\\
THUIR-E-CO-PU-Base-4 		&0.5261 &THUIR-E-CO-PU-Base-4 		&0.7912\\
SLWWW-E-CO-NU-Base-1 	&0.4795 &RUCIR-E-DE-PU-Base-5 		&0.7697\\
RUCIR-E-DE-PU-Base-5 		&0.4315 &SLWWW-E-CO-NU-Base-1 	&0.7152\\
ORG-MANUAL 				&0.4161 &SLWWW-E-CD-NU-Base-3 	&0.7085\\
SLWWW-E-CO-NU-Base-4 	&0.3756 &SLWWW-E-CO-NU-Base-4 	&0.7032\\
SLWWW-E-CD-NU-Base-3 		&0.3748 &ORG-MANUAL 				&0.6519\\
\hline
\end{tabular}

\end{center}
\end{table*}

\begin{table*}[h]
\begin{center}

\caption{Randomised Tukey HSD test results ($B=10,000$ trials) for the corrected WWW-2 results in 
Table~\ref{t:www2corrected-nDCG-Q}.
The runs in the left column statistically significantly outperform those in the right column at the 5\% significance level.
The two-way ANOVA residual variance $V_{\mathrm{E2}}$ for computing effect sizes is 0.0206 for nDCG and 0.0257 for Q.
This table replaces Table~13 of \citet{mao19}.
}\label{t:www2corrected-significance-nDCG-Q} 
\begin{small}
\begin{tabular}{l|l}
\hline
\multicolumn{2}{c}{(a) nDCG}\\
\hline
THUIR-E-CO-MAN-Base-3 	&RUCIR-E-DE-PU-Base-3,RUCIR-E-DE-PU-Base-1,THUIR-E-CO-PU-Base-4,RUCIR-E-DE-PU-Base-5,\\
							&SLWWW-E-CO-NU-Base-1,ORG-MANUAL,SLWWW-E-CO-NU-Base-4,SLWWW-E-CD-NU-Base-3\\
THUIR-E-CO-MAN-Base-2		&RUCIR-E-DE-PU-Base-5,SLWWW-E-CO-NU-Base-1,ORG-MANUAL,SLWWW-E-CO-NU-Base-4,\\
							&SLWWW-E-CD-NU-Base-3\\
THUIR-E-CO-MAN-Base-1 	&\textit{ditto}\\
RUCIR-E-CO-PU-Base-2 		&\textit{ditto}\\
RUCIR-E-DE-PU-Base-4 		&\textit{ditto}\\
MPII-E-CO-NU-Base-2 		&\textit{ditto}\\
MPII-E-CO-NU-Base-1 		&\textit{ditto}\\
MPII-E-CO-NU-Base-5 		&ORG-MANUAL,SLWWW-E-CO-NU-Base-4,SLWWW-E-CD-NU-Base-3\\
MPII-E-CO-NU-Base-3 		&\textit{ditto}\\
baseline\_eng\_v1 				&\textit{ditto}\\
THUIR-E-CO-PU-Base-5 		&\textit{ditto}\\
MPII-E-CO-NU-Base-4 		&\textit{ditto}\\
RUCIR-E-DE-PU-Base-3 		&\textit{ditto}\\
RUCIR-E-DE-PU-Base-1 		&\textit{ditto}\\
THUIR-E-CO-PU-Base-4 		&\textit{ditto}\\
\hline
\multicolumn{2}{c}{(b) Q}\\
\hline
THUIR-E-CO-MAN-Base-3 	&RUCIR-E-DE-PU-Base-3,RUCIR-E-DE-PU-Base-1,RUCIR-E-DE-PU-Base-5,SLWWW-E-CO-NU-Base-1,\\
							&ORG-MANUAL,SLWWW-E-CO-NU-Base-4,SLWWW-E-CD-NU-Base-3\\
THUIR-E-CO-MAN-Base-2 	&RUCIR-E-DE-PU-Base-5,SLWWW-E-CO-NU-Base-1,ORG-MANUAL,SLWWW-E-CO-NU-Base-4,\\
							&SLWWW-E-CD-NU-Base-3\\
THUIR-E-CO-MAN-Base-1 	&\textit{ditto}\\
MPII-E-CO-NU-Base-2 		&\textit{ditto}\\
RUCIR-E-CO-PU-Base-2 		&\textit{ditto}\\
RUCIR-E-DE-PU-Base-4 		&\textit{ditto}\\
MPII-E-CO-NU-Base-1 		&\textit{ditto}\\
MPII-E-CO-NU-Base-5 		&ORG-MANUAL,SLWWW-E-CO-NU-Base-4,SLWWW-E-CD-NU-Base-3\\
MPII-E-CO-NU-Base-3  		&\textit{ditto}\\
baseline\_eng\_v1 				&\textit{ditto}\\
THUIR-E-CO-PU-Base-5 		&\textit{ditto}\\
MPII-E-CO-NU-Base-4 		&\textit{ditto}\\
THUIR-E-CO-PU-Base-4 		&\textit{ditto}\\
RUCIR-E-DE-PU-Base-3 		&\textit{ditto}\\
RUCIR-E-DE-PU-Base-1 		&\textit{ditto}\\
\hline
\end{tabular}
\end{small}
\end{center}
\end{table*}

\begin{table*}[h]
\begin{center}

\caption{Randomised Tukey HSD test results ($B=10,000$ trials) for the corrected WWW-2 results in 
Table~\ref{t:www2corrected-nERR-iRBU}.
The runs in the left column statistically significantly outperform those in the right column at the 5\% significance level.
The two-way ANOVA residual variance $V_{\mathrm{E2}}$ for computing effect sizes is 0.0476 for nERR and 0.0317 for iRBU.
In the uncorrected official results, none of the differences in Mean nERR were statistically significant~\cite[Table~13]{mao19}.
This new table also shows the iRBU results for the sake of consistency 
with our WWW-3 and WWW-4 results.
}\label{t:www2corrected-significance-nERR-iRBU} 
\begin{small}
\begin{tabular}{l|l}
\hline
\multicolumn{2}{c}{(a) nERR}\\
\hline
THUIR-E-CO-MAN-Base-3 	&SLWWW-E-CO-NU-Base-1,RUCIR-E-DE-PU-Base-5,ORG-MANUAL,SLWWW-E-CO-NU-Base-4,\\
							&SLWWW-E-CD-NU-Base-3\\
THUIR-E-CO-MAN-Base-2 	&\textit{ditto}\\
RUCIR-E-CO-PU-Base-2 		&\textit{ditto}\\
RUCIR-E-DE-PU-Base-4 		&RUCIR-E-DE-PU-Base-5,ORG-MANUAL,SLWWW-E-CO-NU-Base-4,SLWWW-E-CD-NU-Base-3\\
THUIR-E-CO-MAN-Base-1 	&\textit{ditto}\\
MPII-E-CO-NU-Base-2 		&\textit{ditto}\\
MPII-E-CO-NU-Base-4 		&\textit{ditto}\\
MPII-E-CO-NU-Base-1 		&\textit{ditto}\\
baseline\_eng\_v1 				&ORG-MANUAL,SLWWW-E-CO-NU-Base-4,SLWWW-E-CD-NU-Base-3\\
THUIR-E-CO-PU-Base-5 		&\textit{ditto}\\
MPII-E-CO-NU-Base-5 		&\textit{ditto}\\
MPII-E-CO-NU-Base-3 		&SLWWW-E-CO-NU-Base-4,SLWWW-E-CD-NU-Base-3\\
RUCIR-E-DE-PU-Base-3 		&\textit{ditto}\\
RUCIR-E-DE-PU-Base-1 		&\textit{ditto}\\
THUIR-E-CO-PU-Base-4 		&\textit{ditto}\\
\hline
\multicolumn{2}{c}{(b) iRBU}\\
\hline
THUIR-E-CO-MAN-Base-3 	&SLWWW-E-CO-NU-Base-1,SLWWW-E-CD-NU-Base-3,SLWWW-E-CO-NU-Base-4,ORG-MANUAL\\
THUIR-E-CO-MAN-Base-2 	&\textit{ditto}\\
RUCIR-E-CO-PU-Base-2 		&\textit{ditto}\\
THUIR-E-CO-MAN-Base-1 	&\textit{ditto}\\
MPII-E-CO-NU-Base-2 		&\textit{ditto}\\
MPII-E-CO-NU-Base-5 		&\textit{ditto}\\
RUCIR-E-DE-PU-Base-4 		&\textit{ditto}\\
MPII-E-CO-NU-Base-3 		&\textit{ditto}\\
RUCIR-E-DE-PU-Base-3 		&\textit{ditto}\\
RUCIR-E-DE-PU-Base-1 		&\textit{ditto}\\
MPII-E-CO-NU-Base-1 		&SLWWW-E-CO-NU-Base-4,ORG-MANUAL\\
baseline\_eng\_v1 				&\textit{ditto}\\
THUIR-E-CO-PU-Base-5 		&\textit{ditto}\\
MPII-E-CO-NU-Base-4 		&ORG-MANUAL\\
THUIR-E-CO-PU-Base-4 		&\textit{ditto}\\
RUCIR-E-DE-PU-Base-5 >		&\textit{ditto}\\
\hline
\end{tabular}
\end{small}
\end{center}
\end{table*}

\begin{table}[t]
\begin{center}
 
\caption{
System ranking similarities for pairs of measures in terms of Kendall's $\tau$ with 95\% CIs ($n=20$ WWW-2 English runs
evaluated with the corrected WWW-2 qrels file).
This table replaces
Table~14 of \citet{mao19}. In addition, iRBU is now included in the comparion.}\label{t:www2corrected-similarities}
\begin{tabular}{c|c|c|c}
\hline
				&Q	&nERR	&iRBU\\
\hline
nDCG	&0.942		&0.858			&0.826\\
			&[0.892, 0.969]&[0.745, 0.923]	&[0.692, 0.905]\\
Q		& -			&0.816			&0.805\\
			&			&[0.676, 0.899]	&[0.658, 0.893]\\
nERR	& -			& -				&0.716\\
			&			&				&[0.519, 0.841]\\
\hline
\end{tabular}

\end{center}
\end{table}

\clearpage

\section{NTCIR-15 English Subtask Results}\label{s:www3}

\begin{table}[t]
\begin{center}
 
\caption{NTCIR-15 English relevance assessment statistics. 
This table replaces Table~5 of \citet{sakai20www3}.
}\label{t:www3assessments}.
\begin{tabular}{c|r|r|r}
\hline
				&WWW-2 topics	&WWW-3 topics		&Total\\
\hline
\#topics			&80				&80			&160\\
\hline
\#assessors/topic	&\multicolumn{3}{c}{8}\\
\hline
Pool depth		&\multicolumn{3}{c}{15}\\
\hline
L4-relevant		&114			&117		&231\\
L3-relevant		&3,893			&3,236		&7,129\\
L2-relevant		&3,872			&4,361		&8,233\\
L1-relevant		&3,942			&3,822		&7,764\\
L0				&3,877			&5,141		&9,018\\
\hline
\#docs pooled		&15,698			&16,677		&32,375\\
\hline
\end{tabular}

\end{center}
\end{table}

This section corrects the English subtask results reported in the WWW-3 overview paper~\cite{sakai20www3}.

The WWW-3 English topic set contains 80 topics (topic IDs: 0101-0180),
just like the WWW-2 English topic set (topic IDs: 0001-0080).
At NTCIR-15, new relevance assessments were collected for all 160 topics,
with eight assessors per topic.
For each topic,
four assessors used a PRI file,
while another four assessors used a RND file.
The official qrels file was formed by
combining the eight 3-point scale relevance labels
by 
taking the integer part of $\log_{2}(S+1)$
as the combined relevance level for each document,
where $S$ is the sum of the eight raw labels.
However, for these assessments collected at NTCIR-15, 
the RND-based relevance labels were incorrect due to the bug.
Hence, every topic was affected by the aforementioned bug.

After correcting the RND-based relevance labels,
we formed a corrected combined qrels file on a 5-point scale.
Table~\ref{t:www3assessments} provides the statistics.
Note that the WWW-2 topic set has 
two versions of (combined) relevance assessments:
the (corrected) WWW-2 qrels file (See Table~\ref{t:www2assessments})
and the (corrected) NTCIR-15 version.

The WWW-3 participants had access to the (noisy) WWW-2 qrels file
before submission,
so their runs were evaluated only on the WWW-3 topic set.
Tables~\ref{t:www3corrected-nDCG-Q}-\ref{t:www3corrected-nERR-iRBU}
show the WWW-3 English run rankings based on (the WWW-3 topic set part of)
 the corrected NTCIR-15 qrels file.
Tables~\ref{t:www3corrected-significance-nDCG}-\ref{t:www3corrected-significance-iRBU} show the corresponding statistical significance test results.
We can observe that:
\begin{itemize}
\item In terms of nDCG, 
mpii-E-CO-NEW-1 and KASYS-E-CO-NEW-\{1,4\} are statistically the best runs:
they statistically significantly outperform 26 other runs (Table~\ref{t:www3corrected-significance-nDCG}).
Similarly, in terms of Q,
mpii-E-CO-NEW-1 and KASYS-E-CO-NEW-\{1,5,4\} are statistically the best runs:
they statistically significantly outperform 25 other runs  (Table~\ref{t:www3corrected-significance-Q}).
\item In terms of nERR,
mpii-E-CO-NEW-1 is statistically the best run:
it is the only run that outperforms 27 other runs (Table~\ref{t:www3corrected-significance-nERR}).
\item In terms of iRBU,
KASYS-E-CO-NEW-4 is  statistically the best run:
it is the only run that outperforms 21 other runs (Table~\ref{t:www3corrected-significance-iRBU}).
\end{itemize}

In the WWW-3 overview paper where the noisy qrels file was used~\cite{sakai20www3},
mpii-E-CO-NEW-1 and KASYS-E-CO-NEW-\{1,4\}
were also statistically the best runs in terms of nDCG and Q,
although KASYS-E-CO-NEW-1 was ranked first in terms of average performance.
The nERR-based results are also similar before and after the bug fix:
mpii-E-CO-NEW-1 is the best run, which suggests
that this run is particularly good at navigational searches,
as was remarked in the overview paper.
The iRBU results are also similar before and after the bug fix:
KASYS-E-CO-NEW-\{4,1\} were statistically the best runs in the overview paper.

For addressing the problem of reproducibility at the WWW-4 task,
we chose KASYS-E-CO-NEW-1 as a state-of-the-art (SOTA) run from the WWW-3 task.
Note that the corrected nDCG and Q results still support this choice,
although mpii-E-CO-NEW-1 would have been a good choice as well.

Table~\ref{t:www3corrected-similarities}
shows the system ranking similarity for each pair of evaluation measure in terms of Kendall's $\tau$,
using (the WWW-3 topic set part of) the corrected NTCIR-15 qrels file
for evaluating the 37 WWW-3 runs.
Again, it can be observed that the four system rankings are quite similar,
with iRBU behaving slightly differently compared to the other three;
the $\tau$'s involving iRBU are noticeably lower than those reported in the NTCIR-15 overview paper.

Table~\ref{t:www3corrected-similarities-www2runs}
shows the system ranking similarity in terms of Kendall's $\tau$,
when the corrected WWW-2 qrels file and (the  WWW-2 topic set part of) the corrected NTCIR-15 qrels file
are compared for each evaluation measure.
It can be observed that the two qrels versions are not interchangeable: they yield somewhat different system rankings.
Similar remarks were made in the overview paper, although the correlations are somewhat higher after the bug fix.

\begin{table*}[t]
\begin{center}
 
\caption{
Corrected results for the WWW-3 runs ($n=80$ WWW-3 topics).
This table replaces the Mean nDCG and Mean Q scores in
Table~11 of \citet{sakai20www3}.}\label{t:www3corrected-nDCG-Q}
\begin{tabular}{l|r||l|r}
\hline
Run							&Mean nDCG@10	&Run				&Mean Q@10\\
\hline
mpii-E-CO-NEW-1 &0.6869 &mpii-E-CO-NEW-1 &0.7009\\
KASYS-E-CO-NEW-1 &0.6866 &KASYS-E-CO-NEW-1 &0.6951\\
KASYS-E-CO-NEW-4 &0.6831 &KASYS-E-CO-NEW-5 &0.6935\\
KASYS-E-CO-NEW-5 &0.6761 &KASYS-E-CO-NEW-4 &0.6921\\
mpii-E-CO-NEW-2 &0.6528 &mpii-E-CO-NEW-2 &0.6680\\
Technion-E-CO-NEW-1 &0.6312 &Technion-E-CO-NEW-1 &0.6559\\
ESTUCeng-E-CO-NEW-3 &0.6296 &ESTUCeng-E-CO-NEW-3 &0.6337\\
ESTUCeng-E-CO-NEW-1 &0.6123 &ESTUCeng-E-CO-NEW-1 &0.6177\\
Technion-E-CO-NEW-4 &0.6000 &Technion-E-CO-NEW-4 &0.6142\\
mpii-E-CO-NEW-3 &0.5992 &mpii-E-CO-NEW-3 &0.6125\\
Technion-E-CO-NEW-2 &0.5856 &Technion-E-CO-NEW-2 &0.6004\\
SLWWW-E-CO-REP-2 &0.5708 &SLWWW-E-CO-REP-2 &0.5921\\
THUIR-E-CO-REV-3 &0.5630 &THUIR-E-CO-REV-3 &0.5757\\
THUIR-E-CO-REV-1 &0.5597 &Technion-E-CO-NEW-5 &0.5698\\
THUIR-E-CO-REV-2 &0.5501 &Technion-E-CO-NEW-3 &0.5608\\
Technion-E-CO-NEW-3 &0.5463 &THUIR-E-CO-REV-1 &0.5588\\
SLWWW-E-CD-NEW-5 &0.5413 &THUIR-E-CO-REV-2 &0.5567\\
KASYS-E-CO-REP-3 &0.5378 &SLWWW-E-CD-NEW-5 &0.5508\\
Technion-E-CO-NEW-5 &0.5371 &KASYS-E-CO-REP-3 &0.5436\\
KASYS-E-CO-REP-2 &0.5332 &KASYS-E-CO-REP-2 &0.5425\\
NAUIR-E-CO-NEW-5 &0.5295 &SLWWW-E-CO-REP-3 &0.5402\\
SLWWW-E-CO-REP-3 &0.5294 &NAUIR-E-CO-NEW-5 &0.5281\\
NAUIR-E-CO-NEW-1 &0.5056 &NAUIR-E-CO-NEW-2 &0.5134\\
NAUIR-E-CO-NEW-2 &0.5049 &NAUIR-E-CO-NEW-1 &0.5105\\
NAUIR-E-CO-NEW-3 &0.5011 &NAUIR-E-CO-NEW-3 &0.5044\\
RUCIR-E-CO-NEW-5 &0.4917 &RUCIR-E-CO-NEW-1 &0.4951\\
NAUIR-E-CO-NEW-4 &0.4860 &RUCIR-E-CO-NEW-5 &0.4911\\
RUCIR-E-CO-NEW-3 &0.4824 &NAUIR-E-CO-NEW-4 &0.4907\\
THUIR-E-CO-NEW-4 &0.4786 &RUCIR-E-CO-NEW-3 &0.4887\\
baselineEng &0.4784 &baselineEng &0.4868\\
SLWWW-E-CO-REP-1 &0.4748 &THUIR-E-CO-NEW-4 &0.4851\\
RUCIR-E-CO-NEW-1 &0.4747 &SLWWW-E-CO-REP-1 &0.4766\\
RUCIR-E-CO-NEW-2 &0.4710 &RUCIR-E-CO-NEW-2 &0.4710\\
ESTUCeng-E-CO-NEW-2 &0.4397 &ESTUCeng-E-CO-NEW-2 &0.4384\\
THUIR-E-CO-REP-5 &0.3913 &THUIR-E-CO-REP-5 &0.4064\\
SLWWW-E-CO-REP-4 &0.3748 &SLWWW-E-CO-REP-4 &0.3688\\
RUCIR-E-CO-NEW-4 &0.3300 &RUCIR-E-CO-NEW-4 &0.3208\\
\hline
\end{tabular}

\end{center}
\end{table*}

\begin{table*}[t]
\begin{center}
 
\caption{
Corrected results for the WWW-3 runs ($n=80$ WWW-3 topics).
This table replaces the Mean nERR and Mean iRBU scores in
Table~12 of \citet{sakai20www3}.}\label{t:www3corrected-nERR-iRBU}
\begin{tabular}{l|r||l|r}
\hline
Run							&Mean nERR@10	&Run				&Mean iRBU@10\\
\hline
mpii-E-CO-NEW-1 &0.8067 &KASYS-E-CO-NEW-4 &0.9526\\
KASYS-E-CO-NEW-1 &0.7895 &KASYS-E-CO-NEW-1 &0.9513\\
KASYS-E-CO-NEW-4 &0.7853 &KASYS-E-CO-NEW-5 &0.9344\\
KASYS-E-CO-NEW-5 &0.7777 &ESTUCeng-E-CO-NEW-3 &0.9199\\
mpii-E-CO-NEW-2 &0.7517 &ESTUCeng-E-CO-NEW-1 &0.9174\\
ESTUCeng-E-CO-NEW-3 &0.7474 &mpii-E-CO-NEW-1 &0.9146\\
Technion-E-CO-NEW-1 &0.7467 &Technion-E-CO-NEW-1 &0.9027\\
ESTUCeng-E-CO-NEW-1 &0.7328 &THUIR-E-CO-REV-1 &0.9005\\
mpii-E-CO-NEW-3 &0.7029 &mpii-E-CO-NEW-2 &0.8944\\
Technion-E-CO-NEW-4 &0.7009 &Technion-E-CO-NEW-2 &0.8900\\
Technion-E-CO-NEW-2 &0.6818 &mpii-E-CO-NEW-3 &0.8861\\
SLWWW-E-CO-REP-2 &0.6747 &THUIR-E-CO-REV-3 &0.8833\\
THUIR-E-CO-REV-2 &0.6719 &THUIR-E-CO-REV-2 &0.8686\\
THUIR-E-CO-REV-1 &0.6633 &NAUIR-E-CO-NEW-5 &0.8646\\
THUIR-E-CO-REV-3 &0.6614 &SLWWW-E-CO-REP-2 &0.8598\\
Technion-E-CO-NEW-3 &0.6539 &THUIR-E-CO-NEW-4 &0.8520\\
KASYS-E-CO-REP-2 &0.6522 &RUCIR-E-CO-NEW-5 &0.8509\\
KASYS-E-CO-REP-3 &0.6489 &Technion-E-CO-NEW-4 &0.8508\\
NAUIR-E-CO-NEW-5 &0.6486 &ESTUCeng-E-CO-NEW-2 &0.8504\\
Technion-E-CO-NEW-5 &0.6432 &SLWWW-E-CO-REP-3 &0.8500\\
SLWWW-E-CO-REP-3 &0.6389 &RUCIR-E-CO-NEW-2 &0.8457\\
SLWWW-E-CD-NEW-5 &0.6288 &SLWWW-E-CO-REP-1 &0.8441\\
NAUIR-E-CO-NEW-4 &0.6211 &NAUIR-E-CO-NEW-4 &0.8431\\
RUCIR-E-CO-NEW-3 &0.6196 &NAUIR-E-CO-NEW-3 &0.8379\\
THUIR-E-CO-NEW-4 &0.6179 &KASYS-E-CO-REP-3 &0.8374\\
NAUIR-E-CO-NEW-2 &0.6065 &Technion-E-CO-NEW-3 &0.8360\\
NAUIR-E-CO-NEW-1 &0.6036 &RUCIR-E-CO-NEW-1 &0.8352\\
NAUIR-E-CO-NEW-3 &0.6031 &KASYS-E-CO-REP-2 &0.8336\\
RUCIR-E-CO-NEW-5 &0.5915 &SLWWW-E-CD-NEW-5 &0.8315\\
RUCIR-E-CO-NEW-1 &0.5849 &RUCIR-E-CO-NEW-3 &0.8294\\
baselineEng &0.5846 &NAUIR-E-CO-NEW-2 &0.8271\\
SLWWW-E-CO-REP-1 &0.5804 &NAUIR-E-CO-NEW-1 &0.8269\\
ESTUCeng-E-CO-NEW-2 &0.5774 &Technion-E-CO-NEW-5 &0.8065\\
RUCIR-E-CO-NEW-2 &0.5712 &baselineEng &0.7991\\
SLWWW-E-CO-REP-4 &0.5080 &SLWWW-E-CO-REP-4 &0.7677\\
THUIR-E-CO-REP-5 &0.5001 &THUIR-E-CO-REP-5 &0.7618\\
RUCIR-E-CO-NEW-4 &0.4431 &RUCIR-E-CO-NEW-4 &0.7452\\
\hline
\end{tabular}

\end{center}
\end{table*}

\begin{table*}[h]
\begin{center}

\caption{Randomised Tukey HSD test results ($B=10,000$ trials) for the corrected WWW-3 results in 
Table~\ref{t:www3corrected-nDCG-Q} (nDCG).
The runs in the left column statistically significantly outperform those in the right column at the 5\% significance level.
The two-way ANOVA residual variance $V_{\mathrm{E2}}$ for computing effect sizes is 
 0.0238
for nDCG.
This table replaces
Table~24(a) of \citet{sakai20www3}.
}\label{t:www3corrected-significance-nDCG} 
\begin{small}
\begin{tabular}{l|l}
\hline
mpii-E-CO-NEW-1 	&SLWWW-E-CO-REP-2,THUIR-E-CO-REV-3,THUIR-E-CO-REV-1,THUIR-E-CO-REV-2,Technion-E-CO-NEW-3,\\
					&SLWWW-E-CD-NEW-5,KASYS-E-CO-REP-3,Technion-E-CO-NEW-5,KASYS-E-CO-REP-2,\\
					&NAUIR-E-CO-NEW-5,SLWWW-E-CO-REP-3,NAUIR-E-CO-NEW-1,NAUIR-E-CO-NEW-2,NAUIR-E-CO-NEW-3,\\
					&RUCIR-E-CO-NEW-5,NAUIR-E-CO-NEW-4,RUCIR-E-CO-NEW-3,THUIR-E-CO-NEW-4,baselineEng,\\
					&SLWWW-E-CO-REP-1,RUCIR-E-CO-NEW-1,RUCIR-E-CO-NEW-2,ESTUCeng-E-CO-NEW-2,\\
					&THUIR-E-CO-REP-5,SLWWW-E-CO-REP-4,RUCIR-E-CO-NEW-4\\
KASYS-E-CO-NEW-1 	&\textit{ditto}\\
KASYS-E-CO-NEW-4 	&\textit{ditto}\\
KASYS-E-CO-NEW-5 	&THUIR-E-CO-REV-3,THUIR-E-CO-REV-1,THUIR-E-CO-REV-2,Technion-E-CO-NEW-3,SLWWW-E-CD-NEW-5,\\
					&KASYS-E-CO-REP-3,Technion-E-CO-NEW-5,KASYS-E-CO-REP-2,NAUIR-E-CO-NEW-5,\\
					&SLWWW-E-CO-REP-3,NAUIR-E-CO-NEW-1,NAUIR-E-CO-NEW-2,NAUIR-E-CO-NEW-3,RUCIR-E-CO-NEW-5,\\
					&NAUIR-E-CO-NEW-4,RUCIR-E-CO-NEW-3,THUIR-E-CO-NEW-4,baselineEng,SLWWW-E-CO-REP-1,\\
					&RUCIR-E-CO-NEW-1,RUCIR-E-CO-NEW-2,ESTUCeng-E-CO-NEW-2,THUIR-E-CO-REP-5,\\
					&SLWWW-E-CO-REP-4,RUCIR-E-CO-NEW-4\\
mpii-E-CO-NEW-2 	&Technion-E-CO-NEW-3,SLWWW-E-CD-NEW-5,KASYS-E-CO-REP-3,Technion-E-CO-NEW-5,\\
					&KASYS-E-CO-REP-2,NAUIR-E-CO-NEW-5,SLWWW-E-CO-REP-3,NAUIR-E-CO-NEW-1,NAUIR-E-CO-NEW-2,\\
					&NAUIR-E-CO-NEW-3,RUCIR-E-CO-NEW-5,NAUIR-E-CO-NEW-4,RUCIR-E-CO-NEW-3,THUIR-E-CO-NEW-4,\\
					&baselineEng,SLWWW-E-CO-REP-1,RUCIR-E-CO-NEW-1,RUCIR-E-CO-NEW-2,ESTUCeng-E-CO-NEW-2,\\
					&THUIR-E-CO-REP-5,SLWWW-E-CO-REP-4,RUCIR-E-CO-NEW-4\\
Technion-E-CO-NEW-1&NAUIR-E-CO-NEW-1,NAUIR-E-CO-NEW-2,NAUIR-E-CO-NEW-3,RUCIR-E-CO-NEW-5,NAUIR-E-CO-NEW-4,\\
					&RUCIR-E-CO-NEW-3,THUIR-E-CO-NEW-4,baselineEng,SLWWW-E-CO-REP-1,RUCIR-E-CO-NEW-1,\\
					&RUCIR-E-CO-NEW-2,ESTUCeng-E-CO-NEW-2,THUIR-E-CO-REP-5,SLWWW-E-CO-REP-4,RUCIR-E-CO-NEW-4\\
ESTUCeng-E-CO-NEW-3	&\textit{ditto}\\
ESTUCeng-E-CO-NEW-1 	&\textit{ditto}\\
Technion-E-CO-NEW-4	&RUCIR-E-CO-NEW-5,NAUIR-E-CO-NEW-4,RUCIR-E-CO-NEW-3,THUIR-E-CO-NEW-4,baselineEng,\\
					&SLWWW-E-CO-REP-1,RUCIR-E-CO-NEW-1,RUCIR-E-CO-NEW-2,ESTUCeng-E-CO-NEW-2,\\
					&THUIR-E-CO-REP-5,SLWWW-E-CO-REP-4,RUCIR-E-CO-NEW-4\\
mpii-E-CO-NEW-3 	&\textit{ditto}\\
Technion-E-CO-NEW-2 &THUIR-E-CO-NEW-4,baselineEng,SLWWW-E-CO-REP-1,RUCIR-E-CO-NEW-1,RUCIR-E-CO-NEW-2,\\
					&ESTUCeng-E-CO-NEW-2,THUIR-E-CO-REP-5,SLWWW-E-CO-REP-4,RUCIR-E-CO-NEW-4\\
SLWWW-E-CO-REP-2 &ESTUCeng-E-CO-NEW-2,THUIR-E-CO-REP-5,SLWWW-E-CO-REP-4,RUCIR-E-CO-NEW-4\\
THUIR-E-CO-REV-3 	&\textit{ditto}\\
THUIR-E-CO-REV-1 	&\textit{ditto}\\
THUIR-E-CO-REV-2 	&\textit{ditto}\\
Technion-E-CO-NEW-3 &\textit{ditto}\\
SLWWW-E-CD-NEW-5 &THUIR-E-CO-REP-5,SLWWW-E-CO-REP-4,RUCIR-E-CO-NEW-4\\
KASYS-E-CO-REP-3 	&\textit{ditto}\\
Technion-E-CO-NEW-5	&\textit{ditto}\\
KASYS-E-CO-REP-2	&\textit{ditto}\\
NAUIR-E-CO-NEW-5	&\textit{ditto}\\
SLWWW-E-CO-REP-3	&\textit{ditto}\\
NAUIR-E-CO-NEW-1	&\textit{ditto}\\
NAUIR-E-CO-NEW-2	&\textit{ditto}\\
NAUIR-E-CO-NEW-3	&\textit{ditto}\\
RUCIR-E-CO-NEW-5	&SLWWW-E-CO-REP-4,RUCIR-E-CO-NEW-4\\
NAUIR-E-CO-NEW-4	&\textit{ditto}\\
RUCIR-E-CO-NEW-3	&\textit{ditto}\\
THUIR-E-CO-NEW-4	&RUCIR-E-CO-NEW-4\\
baselineEng			&\textit{ditto}\\
SLWWW-E-CO-REP-1	&\textit{ditto}\\
RUCIR-E-CO-NEW-1	&\textit{ditto}\\
RUCIR-E-CO-NEW-2	&\textit{ditto}\\
ESTUCeng-E-CO-NEW-2	&\textit{ditto}\\
\hline
\end{tabular}
\end{small}
\end{center}
\end{table*}

\begin{table*}[h]
\begin{center}

\caption{Randomised Tukey HSD test results ($B=10,000$ trials) for the corrected WWW-3 results in 
Table~\ref{t:www3corrected-nDCG-Q} (Q).
The runs in the left column statistically significantly outperform those in the right column at the 5\% significance level.
The two-way ANOVA residual variance $V_{\mathrm{E2}}$ for computing effect sizes is 
0.0284
for Q.
This table replaces
Table~24(b) of \citet{sakai20www3}.
}\label{t:www3corrected-significance-Q} 
\begin{small}
\begin{tabular}{l|l}
\hline
mpii-E-CO-NEW-1	&THUIR-E-CO-REV-3,Technion-E-CO-NEW-5,Technion-E-CO-NEW-3,THUIR-E-CO-REV-1,\\
					&THUIR-E-CO-REV-2,SLWWW-E-CD-NEW-5,KASYS-E-CO-REP-3,KASYS-E-CO-REP-2,SLWWW-E-CO-REP-3,\\
					&NAUIR-E-CO-NEW-5,NAUIR-E-CO-NEW-2,NAUIR-E-CO-NEW-1,NAUIR-E-CO-NEW-3,RUCIR-E-CO-NEW-1,\\
					&RUCIR-E-CO-NEW-5,NAUIR-E-CO-NEW-4,RUCIR-E-CO-NEW-3,baselineEng,THUIR-E-CO-NEW-4,\\
					&SLWWW-E-CO-REP-1,RUCIR-E-CO-NEW-2,ESTUCeng-E-CO-NEW-2,THUIR-E-CO-REP-5,\\
					&SLWWW-E-CO-REP-4,RUCIR-E-CO-NEW-4\\
KASYS-E-CO-NEW-1 	&\textit{ditto}\\ 
KASYS-E-CO-NEW-5 	&\textit{ditto}\\
KASYS-E-CO-NEW-4 	&\textit{ditto}\\
mpii-E-CO-NEW-2 	&SLWWW-E-CD-NEW-5,KASYS-E-CO-REP-3,KASYS-E-CO-REP-2,SLWWW-E-CO-REP-3,NAUIR-E-CO-NEW-5,\\
					&NAUIR-E-CO-NEW-2,NAUIR-E-CO-NEW-1,NAUIR-E-CO-NEW-3,RUCIR-E-CO-NEW-1,RUCIR-E-CO-NEW-5,\\
					&NAUIR-E-CO-NEW-4,RUCIR-E-CO-NEW-3,baselineEng,THUIR-E-CO-NEW-4,SLWWW-E-CO-REP-1,\\
					&RUCIR-E-CO-NEW-2,ESTUCeng-E-CO-NEW-2,THUIR-E-CO-REP-5,SLWWW-E-CO-REP-4,RUCIR-E-CO-NEW-4\\
Technion-E-CO-NEW-1&NAUIR-E-CO-NEW-5,NAUIR-E-CO-NEW-2,NAUIR-E-CO-NEW-1,NAUIR-E-CO-NEW-3,RUCIR-E-CO-NEW-1,\\
					&RUCIR-E-CO-NEW-5,NAUIR-E-CO-NEW-4,RUCIR-E-CO-NEW-3,baselineEng,THUIR-E-CO-NEW-4,\\
					&SLWWW-E-CO-REP-1,RUCIR-E-CO-NEW-2,ESTUCeng-E-CO-NEW-2,THUIR-E-CO-REP-5,\\
					&SLWWW-E-CO-REP-4,RUCIR-E-CO-NEW-4\\
ESTUCeng-E-CO-NEW-3&NAUIR-E-CO-NEW-2,NAUIR-E-CO-NEW-1,NAUIR-E-CO-NEW-3,RUCIR-E-CO-NEW-1,RUCIR-E-CO-NEW-5,\\
					&NAUIR-E-CO-NEW-4,RUCIR-E-CO-NEW-3,baselineEng,THUIR-E-CO-NEW-4,SLWWW-E-CO-REP-1,\\
					&RUCIR-E-CO-NEW-2,ESTUCeng-E-CO-NEW-2,THUIR-E-CO-REP-5,SLWWW-E-CO-REP-4,RUCIR-E-CO-NEW-4\\
ESTUCeng-E-CO-NEW-1	&RUCIR-E-CO-NEW-1,RUCIR-E-CO-NEW-5,NAUIR-E-CO-NEW-4,RUCIR-E-CO-NEW-3,baselineEng,\\
					&THUIR-E-CO-NEW-4,SLWWW-E-CO-REP-1,RUCIR-E-CO-NEW-2,ESTUCeng-E-CO-NEW-2,\\
					&THUIR-E-CO-REP-5,SLWWW-E-CO-REP-4,RUCIR-E-CO-NEW-4\\
Technion-E-CO-NEW-4&\textit{ditto}\\
mpii-E-CO-NEW-3 	&\textit{ditto}\\
Technion-E-CO-NEW-2	&SLWWW-E-CO-REP-1,RUCIR-E-CO-NEW-2,ESTUCeng-E-CO-NEW-2,THUIR-E-CO-REP-5,\\
					&SLWWW-E-CO-REP-4,RUCIR-E-CO-NEW-4\\
SLWWW-E-CO-REP-2 &RUCIR-E-CO-NEW-2,ESTUCeng-E-CO-NEW-2,THUIR-E-CO-REP-5,SLWWW-E-CO-REP-4,RUCIR-E-CO-NEW-4\\
THUIR-E-CO-REV-3 	&ESTUCeng-E-CO-NEW-2,THUIR-E-CO-REP-5,SLWWW-E-CO-REP-4,RUCIR-E-CO-NEW-4\\
Technion-E-CO-NEW-5&\textit{ditto}\\
Technion-E-CO-NEW-3&\textit{ditto}\\
THUIR-E-CO-REV-1 	&\textit{ditto}\\
THUIR-E-CO-REV-2 	&\textit{ditto}\\
SLWWW-E-CD-NEW-5 &THUIR-E-CO-REP-5,SLWWW-E-CO-REP-4,RUCIR-E-CO-NEW-4\\
KASYS-E-CO-REP-3 	&\textit{ditto}\\
KASYS-E-CO-REP-2 	&\textit{ditto}\\
SLWWW-E-CO-REP-3 &\textit{ditto}\\
NAUIR-E-CO-NEW-5 	&\textit{ditto}\\
NAUIR-E-CO-NEW-2 	&SLWWW-E-CO-REP-4,RUCIR-E-CO-NEW-4\\
NAUIR-E-CO-NEW-1 	&\textit{ditto}\\
NAUIR-E-CO-NEW-3 	&\textit{ditto}\\
RUCIR-E-CO-NEW-1 	&\textit{ditto}\\
RUCIR-E-CO-NEW-5 	&\textit{ditto}\\
NAUIR-E-CO-NEW-4 	&\textit{ditto}\\
RUCIR-E-CO-NEW-3 	&\textit{ditto}\\
baselineEng 			&\textit{ditto}\\
THUIR-E-CO-NEW-4 	&\textit{ditto}\\
SLWWW-E-CO-REP-1 &RUCIR-E-CO-NEW-4\\
RUCIR-E-CO-NEW-2 	&\textit{ditto}\\
ESTUCeng-E-CO-NEW-2 &\textit{ditto}\\

\hline
\end{tabular}
\end{small}
\end{center}
\end{table*}

\begin{table*}[h]
\begin{center}

\caption{Randomised Tukey HSD test results ($B=10,000$ trials) for the corrected WWW-3 results in 
Table~\ref{t:www3corrected-nERR-iRBU} (nERR).
The runs in the left column statistically significantly outperform those in the right column at the 5\% significance level.
The two-way ANOVA residual variance $V_{\mathrm{E2}}$ for computing effect sizes is 
 0.0360
for nERR.
This table replaces
Table~25(a) of \citet{sakai20www3}.
}\label{t:www3corrected-significance-nERR} 
\begin{small}
\begin{tabular}{l|l}
\hline
mpii-E-CO-NEW-1 	&Technion-E-CO-NEW-2,SLWWW-E-CO-REP-2,THUIR-E-CO-REV-2,THUIR-E-CO-REV-1,\\
					&THUIR-E-CO-REV-3,Technion-E-CO-NEW-3,KASYS-E-CO-REP-2,KASYS-E-CO-REP-3,\\
					&NAUIR-E-CO-NEW-5,Technion-E-CO-NEW-5,SLWWW-E-CO-REP-3,SLWWW-E-CD-NEW-5,\\
					&NAUIR-E-CO-NEW-4,RUCIR-E-CO-NEW-3,THUIR-E-CO-NEW-4,NAUIR-E-CO-NEW-2,NAUIR-E-CO-NEW-1,\\
					&NAUIR-E-CO-NEW-3,RUCIR-E-CO-NEW-5,RUCIR-E-CO-NEW-1,baselineEng,SLWWW-E-CO-REP-1,\\
					&ESTUCeng-E-CO-NEW-2,RUCIR-E-CO-NEW-2,SLWWW-E-CO-REP-4,THUIR-E-CO-REP-5,RUCIR-E-CO-NEW-4\\
KASYS-E-CO-NEW-1 	&THUIR-E-CO-REV-1,THUIR-E-CO-REV-3,Technion-E-CO-NEW-3,KASYS-E-CO-REP-2,KASYS-E-CO-REP-3,\\
					&NAUIR-E-CO-NEW-5,Technion-E-CO-NEW-5,SLWWW-E-CO-REP-3,SLWWW-E-CD-NEW-5,NAUIR-E-CO-NEW-4,\\
					&RUCIR-E-CO-NEW-3,THUIR-E-CO-NEW-4,NAUIR-E-CO-NEW-2,NAUIR-E-CO-NEW-1,NAUIR-E-CO-NEW-3,\\
					&RUCIR-E-CO-NEW-5,RUCIR-E-CO-NEW-1,baselineEng,SLWWW-E-CO-REP-1,ESTUCeng-E-CO-NEW-2,\\
					&RUCIR-E-CO-NEW-2,SLWWW-E-CO-REP-4,THUIR-E-CO-REP-5,RUCIR-E-CO-NEW-4\\
KASYS-E-CO-NEW-4 	&Technion-E-CO-NEW-3,KASYS-E-CO-REP-2,KASYS-E-CO-REP-3,NAUIR-E-CO-NEW-5,Technion-E-CO-NEW-5,\\
					&SLWWW-E-CO-REP-3,SLWWW-E-CD-NEW-5,NAUIR-E-CO-NEW-4,RUCIR-E-CO-NEW-3,THUIR-E-CO-NEW-4,\\
					&NAUIR-E-CO-NEW-2,NAUIR-E-CO-NEW-1,NAUIR-E-CO-NEW-3,RUCIR-E-CO-NEW-5,RUCIR-E-CO-NEW-1,\\
					&baselineEng,SLWWW-E-CO-REP-1,ESTUCeng-E-CO-NEW-2,RUCIR-E-CO-NEW-2,SLWWW-E-CO-REP-4,\\
					&THUIR-E-CO-REP-5,RUCIR-E-CO-NEW-4\\
KASYS-E-CO-NEW-5 	&KASYS-E-CO-REP-2,KASYS-E-CO-REP-3,NAUIR-E-CO-NEW-5,Technion-E-CO-NEW-5,SLWWW-E-CO-REP-3,\\
					&SLWWW-E-CD-NEW-5,NAUIR-E-CO-NEW-4,RUCIR-E-CO-NEW-3,THUIR-E-CO-NEW-4,NAUIR-E-CO-NEW-2,\\
					&NAUIR-E-CO-NEW-1,NAUIR-E-CO-NEW-3,RUCIR-E-CO-NEW-5,RUCIR-E-CO-NEW-1,baselineEng,\\
					&SLWWW-E-CO-REP-1,ESTUCeng-E-CO-NEW-2,RUCIR-E-CO-NEW-2,SLWWW-E-CO-REP-4,\\
					&THUIR-E-CO-REP-5,RUCIR-E-CO-NEW-4\\
mpii-E-CO-NEW-2 	&NAUIR-E-CO-NEW-4,RUCIR-E-CO-NEW-3,THUIR-E-CO-NEW-4,NAUIR-E-CO-NEW-2,NAUIR-E-CO-NEW-1,\\
					&NAUIR-E-CO-NEW-3,RUCIR-E-CO-NEW-5,RUCIR-E-CO-NEW-1,baselineEng,SLWWW-E-CO-REP-1,\\
					&ESTUCeng-E-CO-NEW-2,RUCIR-E-CO-NEW-2,SLWWW-E-CO-REP-4,THUIR-E-CO-REP-5,RUCIR-E-CO-NEW-4\\
ESTUCeng-E-CO-NEW-3 &\textit{ditto}\\
Technion-E-CO-NEW-1&\textit{ditto}\\
ESTUCeng-E-CO-NEW-1&NAUIR-E-CO-NEW-2,NAUIR-E-CO-NEW-1,NAUIR-E-CO-NEW-3,RUCIR-E-CO-NEW-5,RUCIR-E-CO-NEW-1,\\
					&baselineEng,SLWWW-E-CO-REP-1,ESTUCeng-E-CO-NEW-2,RUCIR-E-CO-NEW-2,SLWWW-E-CO-REP-4,\\
					&THUIR-E-CO-REP-5,RUCIR-E-CO-NEW-4\\
mpii-E-CO-NEW-3 	&ESTUCeng-E-CO-NEW-2,RUCIR-E-CO-NEW-2,SLWWW-E-CO-REP-4,THUIR-E-CO-REP-5,RUCIR-E-CO-NEW-4\\
Technion-E-CO-NEW-4&RUCIR-E-CO-NEW-2,SLWWW-E-CO-REP-4,THUIR-E-CO-REP-5,RUCIR-E-CO-NEW-4\\
Technion-E-CO-NEW-2&SLWWW-E-CO-REP-4,THUIR-E-CO-REP-5,RUCIR-E-CO-NEW-4\\
SLWWW-E-CO-REP-2 &\textit{ditto}\\
THUIR-E-CO-REV-2 	&\textit{ditto}\\
THUIR-E-CO-REV-1 	&\textit{ditto}\\
THUIR-E-CO-REV-3 	&\textit{ditto}\\
Technion-E-CO-NEW-3&\textit{ditto}\\
KASYS-E-CO-REP-2 	&\textit{ditto}\\
KASYS-E-CO-REP-3 	&\textit{ditto}\\
NAUIR-E-CO-NEW-5 	&\textit{ditto}\\
Technion-E-CO-NEW-5&\textit{ditto}\\ 
SLWWW-E-CO-REP-3 &\textit{ditto}\\
SLWWW-E-CD-NEW-5 &THUIR-E-CO-REP-5,RUCIR-E-CO-NEW-4\\
NAUIR-E-CO-NEW-4 	&RUCIR-E-CO-NEW-4\\
RUCIR-E-CO-NEW-3 	&\textit{ditto}\\
THUIR-E-CO-NEW-4 	&\textit{ditto}\\
NAUIR-E-CO-NEW-2 	&\textit{ditto}\\
NAUIR-E-CO-NEW-1 	&\textit{ditto}\\
NAUIR-E-CO-NEW-3 	&\textit{ditto}\\
RUCIR-E-CO-NEW-5 	&\textit{ditto}\\
RUCIR-E-CO-NEW-1 	&\textit{ditto}\\
baselineEng 			&\textit{ditto}\\
SLWWW-E-CO-REP-1	&\textit{ditto}\\
ESTUCeng-E-CO-NEW-2&\textit{ditto}\\
RUCIR-E-CO-NEW-2 	&\textit{ditto}\\
\hline
\end{tabular}
\end{small}
\end{center}
\end{table*}

\begin{table*}[h]
\begin{center}

\caption{Randomised Tukey HSD test results ($B=10,000$ trials) for the corrected WWW-3 results in 
Table~\ref{t:www3corrected-nERR-iRBU} (iRBU).
The runs in the left column statistically significantly outperform those in the right column at the 5\% significance level.
The two-way ANOVA residual variance $V_{\mathrm{E2}}$ for computing effect sizes is 
0.0265
for iRBU.
This table replaces
Table~25(b) of \citet{sakai20www3}.
}\label{t:www3corrected-significance-iRBU} 
\begin{small}
\begin{tabular}{l|l}
\hline
KASYS-E-CO-NEW-4 	&RUCIR-E-CO-NEW-5,Technion-E-CO-NEW-4,ESTUCeng-E-CO-NEW-2,SLWWW-E-CO-REP-3,\\
					&RUCIR-E-CO-NEW-2,SLWWW-E-CO-REP-1,NAUIR-E-CO-NEW-4,NAUIR-E-CO-NEW-3,KASYS-E-CO-REP-3,\\
					&Technion-E-CO-NEW-3,RUCIR-E-CO-NEW-1,KASYS-E-CO-REP-2,SLWWW-E-CD-NEW-5,\\
					&RUCIR-E-CO-NEW-3,NAUIR-E-CO-NEW-2,NAUIR-E-CO-NEW-1,Technion-E-CO-NEW-5,baselineEng,\\
					&SLWWW-E-CO-REP-4,THUIR-E-CO-REP-5,RUCIR-E-CO-NEW-4\\
KASYS-E-CO-NEW-1 	&RUCIR-E-CO-NEW-2,SLWWW-E-CO-REP-1,NAUIR-E-CO-NEW-4,NAUIR-E-CO-NEW-3,KASYS-E-CO-REP-3,\\
					&Technion-E-CO-NEW-3,RUCIR-E-CO-NEW-1,KASYS-E-CO-REP-2,SLWWW-E-CD-NEW-5,RUCIR-E-CO-NEW-3,\\
					&NAUIR-E-CO-NEW-2,NAUIR-E-CO-NEW-1,Technion-E-CO-NEW-5,baselineEng,SLWWW-E-CO-REP-4,\\
					&THUIR-E-CO-REP-5,RUCIR-E-CO-NEW-4\\
KASYS-E-CO-NEW-5 	&SLWWW-E-CD-NEW-5,RUCIR-E-CO-NEW-3,NAUIR-E-CO-NEW-2,NAUIR-E-CO-NEW-1,Technion-E-CO-NEW-5,\\
					&baselineEng,SLWWW-E-CO-REP-4,THUIR-E-CO-REP-5,RUCIR-E-CO-NEW-4\\
ESTUCeng-E-CO-NEW-3	&Technion-E-CO-NEW-5,baselineEng,SLWWW-E-CO-REP-4,THUIR-E-CO-REP-5,RUCIR-E-CO-NEW-4\\
ESTUCeng-E-CO-NEW-1 	&\textit{ditto}\\
mpii-E-CO-NEW-1 	&\textit{ditto}\\
Technion-E-CO-NEW-1	&baselineEng,SLWWW-E-CO-REP-4,THUIR-E-CO-REP-5,RUCIR-E-CO-NEW-4\\
THUIR-E-CO-REV-1 	&SLWWW-E-CO-REP-4,THUIR-E-CO-REP-5,RUCIR-E-CO-NEW-4\\
mpii-E-CO-NEW-2 	&\textit{ditto}\\
Technion-E-CO-NEW-2&\textit{ditto}\\
mpii-E-CO-NEW-3 	&\textit{ditto}\\
THUIR-E-CO-REV-3 	&\textit{ditto}\\
THUIR-E-CO-REV-2 	&THUIR-E-CO-REP-5,RUCIR-E-CO-NEW-4\\
NAUIR-E-CO-NEW-5 	&\textit{ditto}\\
SLWWW-E-CO-REP-2 &RUCIR-E-CO-NEW-4\\
THUIR-E-CO-NEW-4 	&\textit{ditto}\\
RUCIR-E-CO-NEW-5 	&\textit{ditto}\\
Technion-E-CO-NEW-4&\textit{ditto}\\
ESTUCeng-E-CO-NEW-2&\textit{ditto}\\
SLWWW-E-CO-REP-3	&\textit{ditto}\\
\hline
\end{tabular}
\end{small}
\end{center}
\end{table*}

\begin{table}[t]
\begin{center}
 
\caption{
System ranking similarities for pairs of measures in terms of Kendall's $\tau$ with 95\% CIs ($n=37$ WWW-3 English runs
evaluated with the corrected NTCIR-15 qrels file).
This table replaces
Table~14 of \citet{sakai20www3}. 
}\label{t:www3corrected-similarities}
\begin{tabular}{c|c|c|c}
\hline
				&Q	&nERR	&iRBU\\
\hline
nDCG	&0.949		&0.910		&0.604\\
			&[0.921, 0.967]&[0.862, 0.942]&[0.441, 0.728]\\
Q		& -			&0.883		&0.571\\
			&			&[0.822, 0.924]&[0.400, 0.704]\\
nERR	& -			& -			&0.646\\
			&			&			&[0.495, 0.759]\\
\hline
\end{tabular}

\caption{
System ranking similarities for each evaluation measure in terms of Kendall's $\tau$ with 95\% CIs
($n=20$ WWW-2 runs evaluated with the corrected WWW-2 qrels file
and with the corrected NTCIR-15 qrels file).
This table replaces
Table~15 of \citet{sakai20www3}. 
}\label{t:www3corrected-similarities-www2runs}
\begin{tabular}{c|c|c}
\hline
Measure		&$\tau$		&95\%CI\\
\hline
nDCG		&0.742		&[0.559, 0.856]\\
Q			&0.689		&[0.479, 0.824]\\
nERR		&0.747		&[0.566, 0,859]\\
iRBU		&0.653		&[0.427, 0.802]\\
\hline
\end{tabular}

\end{center}
\end{table}

\clearpage

\section{NTCIR-16 WWW-4 Results}\label{s:www4}

This section corrects the results reported in the WWW-4 overview paper~\cite{sakai22www4}.

Unlike the previous WWW test collections which relied on ClueWeb12-B13\footnote{
\url{https://www.lemurproject.org/clueweb12/}
},
the NTCIR-16 WWW-4 test collection introduced a new corpus called Chuweb21.
The WWW-4 test collection has 50 topics (0201-0250).
Each topic was judged independently by three assessors:
a Gold assessor (i.e., topic creator),
a Bronze-Waseda (BronzeW) assessor,
and
a Bronze-Tsinghua (BronzeT) assessor.
Seven of the WWW-4 task organisers served as Gold assessors.
BronzeW assessors are international course students of 
Waseda University, Japan; previous WWW English subtasks
relied entirely on BronzeW assessors.
BronzeT assessors are professional labellers working for a Chinese company.

The Bronze-All qrels file used in the WWW-4 overview paper
was created by
simply summing the 3-point scale BronzeW and BronzeT relevance labels;
hence the qrels file is on a 5-point scale (L0-L4).
The Bronze-All-based results in the overview paper are correct,
because all Bronze assessors used a PRI file for each topic,
and all PRI-based assessments for the WWW-4 test collection happened to be correct.
On the other hand, the Gold-based results in the overview paper are incorrect,
as the Gold qrels file suffers from the aforementioned bug.
More specifically,
the Gold assessors used a PRI file for 25 topics,\footnote{
0201, 0204, 0205, 0208, 0209, 0212, 0213, 0214, 0217, 0218, 0219, 0222, 0223, 0225, 0228, 0229, 0231, 0235, 0236, 0239, 0242, 0245, 0246, 0248, 0249
} and a RND file for the remaining 25 topics,\footnote{
0202, 0203, 0206, 0207, 0210, 0211, 0215, 0216, 0220, 0221, 0224, 0226, 0227, 0230, 0232, 0233, 0234, 0237, 0238, 0240, 0241, 0243, 0244, 0247, 0250
}
and the RND-based relevance assessments are incorrect.
Hereafter, we report on results based on the corrected Gold qrels file.

\begin{table*}[t]
\begin{center}
 
\caption{NTCIR-14 WWW-2 English relevance assessment statistics.
This is identical to Table~4 from \citet{sakai22www4}:
although the original Gold qrels file contained noise, 
the distribution of L0-L4 labels was correct.
It was the document-label mapping that was incorrect.
}\label{t:www4assessments}

\begin{tabular}{c|r|rrr}
\hline
relevance 	&Gold				&Bronze-Waseda		&Bronze-Tsinghua		&Bronze-All\\
level			&(1 assessor/topic)	&(1 assessor/topic)	&(1 assessor/topic)	&(2 assessors/topic)\\
\hline
L0			&7,154				&5,584				&6,571				&4,900\\
L1			&1,806				&3,158				&1,986				&1,881\\
L2			&1,373				&1,591				&1,776				&1,485\\
L3			&N/A				&N/A				&N/A				&1,241\\
L4			&N/A				&N/A				&N/A				&826\\
\hline
total			&10,333				&10,333				&10,333				&10,333\\
\hline
\end{tabular}

\end{center}
\end{table*}

Table~\ref{t:www4assessments}  shows the statistics of the corrected Gold WWW-4 qrels file,
together with the Bronze statistics for comparison.
The table is actually the same as the one in the overview paper:
the label distribution for the Gold qrels file remains unchanged;
only the document-label mapping was changed after the bug fix.

\begin{table}[t]
\begin{center}
 
\caption{Mean per-topic inter-assessor agreement in terms of quadratic weighted Cohen's $\kappa$ ($n=50$ topics).
This table replaces Table~5 of \citet{sakai22www4}; 
the Waseda-Tsinghua agreement is unchanged as this comparison does not involve Gold assessments that contained noise.}\label{t:inter-assessor}

\begin{tabular}{c|r}
\hline
qrels version		&mean $\kappa$\\
\hline
Gold-Waseda		&0.440\\
Gold-Tsinghua	&0.495\\
Waseda-Tsinghua	&0.458\\
\hline
\end{tabular}

\caption{Comparison of the mean $\kappa$'s with a randomised Tukey HSD test ($B=5,000$ trials). The effect sizes are based on the two-way ANOVA residual variance 
$V_{E2}=0.0173$~\cite{sakai18book}.
This table replaces Table~6 of \citet{sakai22www4}.
}\label{t:inter-assessor-significance}

\begin{tabular}{c|r}
\hline
Gold-Waseda vs. Gold-Tsinghua		&$p=0.097, \textit{ES}_{E2}=0.417$\\
Gold-Waseda vs. Waseda-Tsinghua	&$p=0.781, \textit{ES}_{E2}=0.135$\\
Gold-Tsinghua vs. Waseda-Tsinghua 	&$p=0.354, \textit{ES}_{E2}=0.282$\\
\hline
\end{tabular}

\end{center}
\end{table}

Table~\ref{t:inter-assessor} shows the mean inter-assessor agreement 
for each pair of qrels files, after fixing the Gold qrels file.
Before the bug fix, the mean Gold-Waseda and Gold-Tsinghua 
agreements were 0.242 and 0.280, respectively;
it can be observed that the mean agreements 
after the bug fix are reasonable,
and also similar to the mean Waseda-Tsinghua agreement
(which is unaffected by the bug since it is independent of 
 the Gold assessments).
Table~\ref{t:inter-assessor-significance}
compares the mean $\kappa$'s shown in 
Table~\ref{t:inter-assessor} in terms of statistical significance.
It can be observed that the differences in mean inter-assessor agreements
are not statistically significant.

Tables~\ref{t:per-assessor-gold} and~\ref{t:per-assessor-bronze}
examine the per-topic $\kappa$'s at the individual assessor level,
after fixing the Gold qrels file.
For example, the ``Gold01'' row of Table~\ref{t:per-assessor-gold}
compares the labels of Gold01 (the first author of this paper)
with those of Waseda and Tsinghua assessors,
and shows the mean $\kappa$ over the eight topics that she was in charge of.
Before the bug fix, the mean $\kappa$'s were in the range of 0.145-0.507,
whereas the new mean $\kappa$'s are in the range of 0.329-0.623.
In short, the corrected mean inter-assessor agreements look ``normal.''




\begin{table}[t]
\begin{center}

\caption{Mean per-topic inter-assessor agreement for each gold assessor in terms of quadratic weighted Cohen's $\kappa$
 ($n=8$ topics for Gold01; $n=7$ topics for the others). For example, the labels of Gold01 are compared with those given by the Waseda and Tsinghua bronze assessors.
This table replaces Table~7 of \citet{sakai22www4}. 
}\label{t:per-assessor-gold}

\begin{tabular}{c|rr}
\hline
assessor&mean $\kappa$ (with Waseda)	&mean $\kappa$ (with Tsinghua)\\
\hline
Gold01	&0.393	&0.473\\
Gold02	&0.438	&0.583\\
Gold03	&0.329	&0.366\\
Gold04	&0.504	&0.473\\
Gold05	&0.453	&0.486\\
Gold06	&0.414	&0.463\\
Gold07	&0.554	&0.623\\
\hline
\end{tabular}
 
\caption{Mean per-topic inter-assessor agreement for each bronze assessor in terms of quadratic weighted Cohen's $\kappa$
 ($n=10$ topics). For example, the labels of Waseda01 are compared with those given by the Gold and Tsinghua assessors.
This table replaces Table~8 of \citet{sakai22www4}.
The values in the rightmost column are unchanged
as they do not involve Gold assessments.
}\label{t:per-assessor-bronze}

\begin{tabular}{c|rr}
\hline
assessor&mean $\kappa$ (with Gold)	&mean $\kappa$ (with Tsinghua)\\
\hline
Waseda01	&0.395	&0.450\\
Waseda02	&0.460	&0.459\\
Waseda03	&0.490	&0.444\\
Waseda04	&0.426	&0.428\\
Waseda05	&0.428	&0.507\\
\hline
assessor&mean $\kappa$ (with Gold)	&mean $\kappa$ (with Waseda)\\
\hline
Tsinghua06	&0.549	&0.476\\
Tsinghua07	&0.480	&0.485\\
Tsinghua08	&0.462	&0.395\\
Tsinghua09	&0.481	&0.500\\
Tsinghua10	&0.501	&0.432\\
\hline
\end{tabular}

\end{center}
\end{table}

Table~\ref{t:results-gold} shows the 
WWW-4 run rankings based on the corrected Gold qrels file.
Table~\ref{t:gold-significance} shows the corresponding statistical significance test results.
The following observations can be made.
\begin{itemize}
\item In terms of nDCG, 
THUIR-CO-NEW-2
is statistically the best run,
in that it is the only run that statistically significantly outperforms eight other runs.
Q also agrees, although both 
THUIR-CO-NEW-\{2,1\} statistically significantly
outperform nine other runs.
However, even THUIR-CO-NEW-2
is statistically indistinguishable from \textbf{KASYS-CO-REV-6}, the SOTA from WWW-3
(equivalent to the WWW-3 run KASYS-E-CO-NEW-1).
Also, note that \textbf{KASYS-CO-REV-6} is ranked first in the Mean iRBU ranking.
Hence, we cannot conclude from these results that the improvement over the WWW-3 SOTA is substantial.
\item In terms of nERR, which is a measure suitable for navigational search,
ORG-TOPICDEV is the top run: this is the only run that outperforms seven other runs.
This is not surprising, as it contains only the seed relevant documents identified 
by the Gold assessors at topic development time.
\item SLWWW-CO-REP-1, which is our only REP run,
performs very similarly to \textbf{KASYS-CO-REV-6}
in terms of all four evaluation measures.
This suggests that the reproducibility effort may be successful to some degree.
\end{itemize}

In the WWW-4 overview paper where the noisy Gold qrels file was used~\cite{sakai22www4},
the statistical significance test results were highly inconclusive:
in terms of nDCG, Q, and iRBU,
17 out of the 18 runs were statistically tied:
they all statistically significantly outperformed ORG-TOPICDEV.
No statistically significant differences were found with nERR.
In contrast, our corrected results show that
THUIR-CO-NEW-2 does better than some of the other runs,
which is consistent with the Bronze-All results
reported in the WWW-4 overview paper.
Also, the nERR-based rankings are strikingly different 
before and after the bug fix:
ORG-TOPICDEV, the top performer in the corrected results,
was ranked near the bottom in the overview paper.
This is probably because the relevant documents identified by the Gold assessors 
at topic development time
were not labelled as relevant in the Gold qrels file due to the bug.

\begin{table*}[t]
\begin{center}
 
\caption{Results for the WWW-4 runs based on the corrected Gold qrels file
($n=50$ WWW-4 test topics).
This table replaces Table~9 of \citet{sakai22www4}.
}\label{t:results-gold}

\begin{tabular}{lr|lr}
\hline
Run name	&(a) Mean nDCG	&Run name 	&(b) Mean Q\\
\hline
%
THUIR-CO-NEW-2 &0.5157 &THUIR-CO-NEW-2 &0.4491\\
THUIR-CO-NEW-1 &0.4994 &THUIR-CO-NEW-1 &0.4434\\
\textbf{KASYS-CO-REV-6} &0.4855 &SLWWW-CO-NEW-4 &0.3994\\
SLWWW-CO-REP-1 &0.4833 &\textbf{KASYS-CO-REV-6} &0.3911\\
SLWWW-CO-NEW-4 &0.4708 &SLWWW-CO-REP-1 &0.3891\\
SLWWW-CO-NEW-2 &0.4578 &SLWWW-CO-NEW-2 &0.3884\\
THUIR-CO-NEW-3 &0.4416 &SLWWW-CO-NEW-3 &0.3682\\
SLWWW-CO-NEW-3 &0.4353 &THUIR-CO-NEW-3 &0.3659\\
SLWWW-CO-NEW-5 &0.4211 &SLWWW-CO-NEW-5 &0.3515\\
THUIR-CO-NEW-5 &0.4069 &THUIR-CO-NEW-5 &0.3250\\
KASYS-CO-NEW-4 &0.3859 &KASYS-CO-NEW-4 &0.3184\\
KASYS-CO-NEW-2 &0.3858 &KASYS-CD-NEW-1 &0.3169\\
KASYS-CD-NEW-1 &0.3852 &KASYS-CO-NEW-2 &0.3164\\
baseline &0.3824 &KASYS-CD-NEW-3 &0.3154\\
THUIR-CO-NEW-4 &0.3820 &baseline &0.3126\\
KASYS-CD-NEW-3 &0.3812 &THUIR-CO-NEW-4 &0.3013\\
KASYS-CD-NEW-5 &0.3107 &KASYS-CD-NEW-5 &0.2350\\
ORG-TOPICDEV &0.2802 &ORG-TOPICDEV &0.1587\\

\hline
Run name	&(c) Mean nERR	&Run name 	&(d) Mean iRBU\\
\hline
%
ORG-TOPICDEV &0.7213 &\textbf{KASYS-CO-REV-6} &0.8846\\
THUIR-CO-NEW-2 &0.6983 &SLWWW-CO-REP-1 &0.8842\\
THUIR-CO-NEW-1 &0.6753 &THUIR-CO-NEW-2 &0.8538\\
\textbf{KASYS-CO-REV-6} &0.6554 &SLWWW-CO-NEW-4 &0.8378\\
SLWWW-CO-REP-1 &0.6481 &SLWWW-CO-NEW-2 &0.8333\\
SLWWW-CO-NEW-2 &0.6214 &THUIR-CO-NEW-1 &0.8307\\
SLWWW-CO-NEW-4 &0.6115 &THUIR-CO-NEW-3 &0.8149\\
SLWWW-CO-NEW-3 &0.5986 &SLWWW-CO-NEW-3 &0.8114\\
THUIR-CO-NEW-3 &0.5946 &THUIR-CO-NEW-5 &0.8109\\
SLWWW-CO-NEW-5 &0.5720 &SLWWW-CO-NEW-5 &0.7848\\
THUIR-CO-NEW-4 &0.5578 &THUIR-CO-NEW-4 &0.7627\\
KASYS-CO-NEW-4 &0.5386 &KASYS-CD-NEW-1 &0.7504\\
THUIR-CO-NEW-5 &0.5373 &KASYS-CO-NEW-2 &0.7500\\
baseline &0.5155 &KASYS-CD-NEW-3 &0.7500\\
KASYS-CO-NEW-2 &0.5145 &KASYS-CO-NEW-4 &0.7498\\
KASYS-CD-NEW-1 &0.5058 &baseline &0.7497\\
KASYS-CD-NEW-3 &0.5033 &KASYS-CD-NEW-5 &0.7321\\
KASYS-CD-NEW-5 &0.4760 &ORG-TOPICDEV &0.6794\\
\hline
\end{tabular}

\end{center}
\end{table*}

\begin{table*}[h]
\begin{center}

\caption{Randomised Tukey HSD test results ($B=5,000$ trials) for the Gold-based results in Table~\ref{t:results-gold}.
The runs in the left column statistically significantly outperform those in the right column at the 5\% significance level.
This table replaces Table~10 of \citet{sakai22www4}.
}\label{t:gold-significance}
\begin{small}
\begin{tabular}{l|l}
\hline
\multicolumn{2}{c}{(a) nDCG}\\
\hline
THUIR-CO-NEW-2 	&KASYS-CO-NEW-4,KASYS-CO-NEW-2,KASYS-CD-NEW-1,baseline,THUIR-CO-NEW-4,\\
					&KASYS-CD-NEW-3,KASYS-CD-NEW-5,ORG-TOPICDEV\\
THUIR-CO-NEW-1 	&KASYS-CD-NEW-1,baseline,THUIR-CO-NEW-4,KASYS-CD-NEW-3,KASYS-CD-NEW-5,ORG-TOPICDEV\\
KASYS-CO-REV-6 	&KASYS-CD-NEW-5,ORG-TOPICDEV\\
SLWWW-CO-REP-1 	&\textit{ditto}\\
SLWWW-CO-NEW-4 	&\textit{ditto}\\
SLWWW-CO-NEW-2 	&\textit{ditto}\\
THUIR-CO-NEW-3 	&\textit{ditto}\\
SLWWW-CO-NEW-3 	&\textit{ditto}\\
SLWWW-CO-NEW-5 	&ORG-TOPICDEV\\
THUIR-CO-NEW-5 	&\textit{ditto}\\
\hline
\multicolumn{2}{c}{(b) Q}\\
\hline
THUIR-CO-NEW-2 	&THUIR-CO-NEW-5,KASYS-CO-NEW-4,KASYS-CD-NEW-1,KASYS-CO-NEW-2,\\
					&KASYS-CD-NEW-3,baseline,THUIR-CO-NEW-4,KASYS-CD-NEW-5,ORG-TOPICDEV\\
THUIR-CO-NEW-1 	&\textit{ditto}\\
SLWWW-CO-NEW-4 	&KASYS-CD-NEW-5,ORG-TOPICDEV\\
KASYS-CO-REV-6 	&\textit{ditto}\\
SLWWW-CO-REP-1 	&\textit{ditto}\\
SLWWW-CO-NEW-2 	&\textit{ditto}\\
SLWWW-CO-NEW-3 	&\textit{ditto}\\
THUIR-CO-NEW-3 	&\textit{ditto}\\
SLWWW-CO-NEW-5 	&ORG-TOPICDEV\\
THUIR-CO-NEW-5 	&\textit{ditto}\\
KASYS-CO-NEW-4 	&\textit{ditto}\\
KASYS-CD-NEW-1 	&\textit{ditto}\\
KASYS-CO-NEW-2 	&\textit{ditto}\\
KASYS-CD-NEW-3 	&\textit{ditto}\\
baseline 				&\textit{ditto}\\
THUIR-CO-NEW-4 	&\textit{ditto}\\
\hline
\multicolumn{2}{c}{(c) nERR}\\
\hline
ORG-TOPICDEV 		&KASYS-CO-NEW-4,THUIR-CO-NEW-5,baseline,KASYS-CO-NEW-2,KASYS-CD-NEW-1,\\
					&KASYS-CD-NEW-3,KASYS-CD-NEW-5\\
THUIR-CO-NEW-2 	&baseline,KASYS-CO-NEW-2,KASYS-CD-NEW-1,KASYS-CD-NEW-3,KASYS-CD-NEW-5\\
THUIR-CO-NEW-1 	&KASYS-CD-NEW-3,KASYS-CD-NEW-5\\
KASYS-CO-REV-6 	&KASYS-CD-NEW-5\\
SLWWW-CO-REP-1 	&\textit{ditto}\\
\hline
\multicolumn{2}{c}{(d) iRBU}\\
\hline
KASYS-CO-REV-6 	&KASYS-CD-NEW-5,ORG-TOPICDEV\\
SLWWW-CO-REP-1 	&\textit{ditto}\\
THUIR-CO-NEW-2 	&ORG-TOPICDEV\\
SLWWW-CO-NEW-4 	&\textit{ditto}\\
SLWWW-CO-NEW-2 	&\textit{ditto}\\
THUIR-CO-NEW-1 	&\textit{ditto}\\
\hline
\end{tabular}
\end{small}

\end{center}
\end{table*}

\begin{table*}[t]
\begin{center}

\caption{Corrected run ranking correlations in terms of Kendall's $\tau$ with 95\%CIs ($n=18$ runs).
This table replaces Table~14 of \citet{sakai22www4};
Part~(b) is unchanged as it does not rely on the Gold qrels file.
}\label{t:www4run-ranking-correlation}

\begin{tabular}{c|c|c|c}
\hline
(a) Gold		&Q						&nERR				&iRBU\\
\hline
nDCG		&0.922 [0.850, 0.960]		&0.647 [0.400, 0.806]	&0.784 [0.610, 0.886]\\
Q			& - 						&0.595 [0.327, 0.775]	&0.758 [0.610, 0.886]\\
nERR		& - 						& - 					&0.536 [0.247, 0.737]\\
\hline
(b) Bronze-All	&Q						&nERR				&iRBU\\
\hline
nDCG		&0.961 [0.924, 0.980]		&0.725 [0.517, 0.852]	&0.699 [0.477, 0.837]\\
Q			& -						&0.686 [0.457, 0.830]	&0.712 [0.497, 0.845]\\
nERR		& - 						& -					&0.503 [0.204, 0.716]\\
\hline
\end{tabular}

\begin{tabular}{c|l}
\multicolumn{2}{c}{(c) Gold vs. Bronze-All}\\
\hline
nDCG	&0.804 [0.643, 0.897]\\
Q		&0.791 [0.622, 0.890]\\
nERR	&0.752 [0.559, 0.868]\\
iRBU	&0.614 [0.353, 0.786]\\
\hline
\end{tabular}

\end{center}
\end{table*}

Table~\ref{t:www4run-ranking-correlation} Part~(a)
compares the system rankings according to the four evaluation measures 
with the corrected qrels file;
Part~(b) does the same with the Bronze-All qrels file (unchanged from the overview paper);
Part~(c) compares the Gold and Bronze-All run rankings when the same evaluation measure is used.
Part~(c) is substantially different after the bug fix:
in the WWW-4 overview paper, the Gold-vs-Bronze-all $\tau$'s were in the range of 0.327-0.680;
in contrast, 
our new $\tau$'s are in the range of 0.614-0.804.
This means that the original Gold-based rankings were quite wrong.

\clearpage

\section{Reproducibility Results}

We evaluate reproducibility with the measures defined in~\cite{breuer20}.
All measures are computed with the repro{\_}eval library~\cite{breuer21ecir}.
In WWW-3 we have both reproducibility and replicability results, in WWW-4 we only have reproducibility results. 

\subsection{WWW-3 Reproducibility Results}

This section corrects the reproducibility results reported in the WWW-3 overview paper~\cite{sakai20www3}.
Note that WWW-3 overview paper uses an old version of the ACM Artifact Review and Badging Policy\footnote{https://www.acm.org/publications/policies/artifact-review-badging}. 
In this paper we use the latest version instead\footnote{https://www.acm.org/publications/policies/artifact-review-and-badging-current}, which swaps the meaning of the two terms reproducibility and replicability.
Therefore, reproducibility and replicability are swapped when comparing this section with WWW-3 overview paper.

The targets of the reproducibility experiment are $2$ runs submitted at WWW-2: one advanced A-run (THUIR-E-CO-MAN-Base2) and one baseline B-run (THUIR-E-CO-PU-Base4).
We evaluate $2$ REP A-runs (KASYS-E-CO-REP-2 and SLWWW-E-CO-REP-4) and $1$ REP B-run (KASYS-E-CO-REP-3). 
For reproducibility, we re-evaluate both the original and reproduced runs with the corrected qrels from WWW-2. 
For replicability, we use the corrected WWW-2 qrels for the original runs with WWW-2 topics and the corrected WWW-3 qrels for replicated runs with WWW-3 topics.

\subsubsection{Reproducibility}

Results reporting the ordering of documents (Table 16 in~\cite{sakai20www3}) compare the original and reproduced runs with Kendall's $\tau$ union and Rank Biased Overlap (RBO) and do not use the qrels.
Therefore, these results are not affected by the aforementioned bug.

\begin{table}[tb]
\caption{$\text{RMSE}_{abs}$ scores for each replicability runs and measures. This table replaces Table~17 of~\citet{sakai20www3}.}
\label{tab:centre_rmse_abs}
\resizebox{\columnwidth}{!}{%
\begin{tabular}{llllll}
\toprule
          &                      & \multicolumn{4}{c}{$\text{RMSE}_{abs}$} \\
Run Type  & Run Name             & nDCG     & Q        & nERR     & iRBU    \\
\midrule
REP A-run & KASYS-E-CO-REP-2 & 0.2003   & 0.2288   & 0.3047  & 0.2352  \\
REP A-run & SLWWW-E-CO-REP-4 & 0.3686   & 0.4192   & 0.4977  & 0.4794  \\
REP B-run & KASYS-E-CO-REP-3 & 0.2567   & 0.2739   & 0.3622  & 0.3591  \\
\bottomrule
\end{tabular}
}
\end{table}

Table~\ref{tab:centre_rmse_abs} reports the corrected $\text{RMSE}_{abs}$ scores (the lower the better). 
Overall $\text{RMSE}_{abs}$ are slightly higher when computed with the corrected qrels.
This strengthens the conclusion from~\cite{sakai20www3}, that \textit{none of the REP runs could successfully replicate the effectiveness scores of the original runs}.
The trend among runs remains unchanged, with KASYS-E-CO-REP-2 being better than SLWWW-E-CO-REP-4.

\begin{table*}[tb]
\caption{$p$-value returned by a two tailed paired t-test run between the original and reproduced runs. This table replaces Table~18 of~\citet{sakai20www3}.}
\label{tab:centre_pvalue}
\begin{tabular}{@{}llS[table-format = 1.4e-2]S[table-format = 1.4e-2]S[table-format = 1.4e-2]S[table-format = 1.4e-2]@{}}
\toprule
          &                  & \multicolumn{4}{c}{$p$-value}                     \\
Run Type  & Run Name         & \multicolumn{1}{l}{nDCG}       & \multicolumn{1}{l}{Q}          & \multicolumn{1}{l}{nERR}        &\multicolumn{1}{l}{iRBU}       \\ \midrule
REP A-run & KASYS-E-CO-REP-2 & 2.1022e-06     & 1.0804e-06 & 2.7970e-04     & 2.0544e-04     \\
REP A-run & SLWWW-E-CO-REP-4 & 7.7170e-26 & 3.7204e-24 & 1.5874e-17 & 2.3279e-16 \\
REP B-run & KASYS-E-CO-REP-3 & 0.7819     & 0.8381     & 0.9364     & 0.3702     \\ \bottomrule
\end{tabular}
\end{table*}

Table~\ref{tab:centre_pvalue} reports the corrected $p$-values returned from a two tailed paired t-test (the lower the $p$-value, the higher the evidence that the original and reproduced runs are significantly different).
For 
REP A-runs,
the corrected $p$-values are smaller, which further supports the conclusion from~\cite{sakai20www3} that 
REP A-runs
are significantly different from the original advanced run.
For 
REP B-runs
the corrected $p$-values are smaller and again this strengthen the observation that the 
REP B-run
is not significantly different from the original baseline run.

\begin{table*}[htb]
\caption{Results for reproducibility of effects over a baseline. This table replaces Table~19 of~\citet{sakai20www3}.}
\label{tab:centre_repli_effect_b}
\resizebox{\textwidth}{!}{%
\begin{tabular}{@{}ll|llll|llll|llll@{}}
\toprule
                 &                  & \multicolumn{4}{c|}{${\it RMSE}_{\Delta}$} & \multicolumn{4}{c|}{${\it ER}_{\it repro}$}              & \multicolumn{4}{c}{$\Delta RI_{\it repro}$}   \\
A-run            & B-run            & nDCG      & Q         & nERR      & iRBU     & nDCG    & Q       & nERR    & iRBU    & nDCG   & Q      & nERR    & iRBU   \\ \midrule
KASYS-E-CO-REP-2 & KASYS-E-CO-REP-3 & 0.2658    & 0.2921    & 0.3924   & 0.3236   & -0.5029 & -0.6020 & -0.0604 & 0.0846 & 0.2846 & 0.3262 & 0.2340 & 0.0732 \\ \bottomrule
\end{tabular}
}
\end{table*}

Finally, Table~\ref{tab:centre_repli_effect_b} shows results of reproducibility of the effect over a baseline.
Scores for $\text{RMSE}_{\Delta}$ slightly increase, as it happens for $\text{RMSE}_{abs}$.
The same happens for $\Delta\text{RI}_{repro}$.
For $\text{ER}_{repro}$, scores get closer to zero, meaning that KASYS REP runs could not reproduce the same effects of the original runs.
Again, this results supports the conclusions from~\citet{sakai20www3}.

\subsubsection{Replicability}

\begin{table*}[tb]
\caption{$p$-value returned by a two tailed unpaired t-test run between the original and replicated runs. This table replaces Table~20 of~\citet{sakai20www3}.}
\label{tab:centre_repro_pvalue}
\begin{tabular}{@{}llS[table-format = 1.4e-2]S[table-format = 1.4e-2]S[table-format = 1.4e-2]S[table-format = 1.4e-2]@{}}
\toprule
          &                  & \multicolumn{4}{c}{$p$-value}                     \\
Run Type  & Run Name         & \multicolumn{1}{l}{nDCG}       & \multicolumn{1}{l}{Q}          & \multicolumn{1}{l}{nERR}        & \multicolumn{1}{l}{iRBU}       \\ \midrule
REP A-run & KASYS-E-CO-REP-2 & 0.0662     & 0.0431 & 0.8180     & 0.6015     \\
REP A-run & SLWWW-E-CO-REP-4 & 0.0225         & 0.0526     & 2.8406e-03        & 0.0296 \\
REP B-run & KASYS-E-CO-REP-3 & 1.9769e-04     & 1.8240e-04     & 5.3593e-03     & 0.3011     \\ \bottomrule
\end{tabular}
\end{table*}

Table~\ref{tab:centre_repro_pvalue} reports the corrected $p$-values returned from a two tailed unpaired t-test (the lower the $p$-value, the higher the evidence that the original and replicated runs are significantly different).
Results are somehow different from those reported in~\cite{sakai20www3}, overall all $p$-values increased and:
\begin{itemize}
\item KASYS-E-CO-REP-2 is not significantly different from the original run, but it was in~\cite{sakai20www3};
\item SLWWW-E-CO-REP-4 is not significantly different from the original run with Q measure, but it was in~\cite{sakai20www3}, and it is significantly different from the original run with nERR and iRBU, but it was not in~\cite{sakai20www3}.
\item KASYS-E-CO-REP-3 is not significantly different from the original run with iRBU, but it was in~\cite{sakai20www3}.
\end{itemize}
These results suggest that among 
REP A-runs,
KASYS-E-CO-REP-2 replicates better the original run, which is now consistent with the reproducibility results in Table~\ref{tab:centre_pvalue}.
Similarly, the increased $p$-values for the 
REP B-run
better align with the reproducibility results.

\begin{table*}[tb]
\caption{Results for replicability of effects over a baseline. This table replaces Table~21 of~\citet{sakai20www3}.}
\label{tab:centre_repro_effect_b}
\begin{tabular}{@{}ll|llll|llll@{}}
\toprule
                 &                   & \multicolumn{4}{c|}{${\it ER}_{\it repli}$}              & \multicolumn{4}{c}{$\Delta RI_{\it repli}$}   \\
A-run            & B-run            & nDCG    & Q       & nERR    & iRBU    & nDCG   & Q      & nERR    & iRBU   \\ \midrule
KASYS-E-CO-REP-2 & KASYS-E-CO-REP-3   & -0.0617 & -0.0140 & 0.0290 & -0.0601
& 0.1991 & 0.2069 & 0.2156 & 0.0849 \\ \bottomrule
\end{tabular}
\end{table*}

Finally, Table~\ref{tab:centre_repro_effect_b} shows results of replicability of the effect over a baseline.
The corrected replicability results exhibit a similar trend to the corrected reproducibility results: $\text{ER}_{\mathrm{repli}}$ scores get closer to zero and $\Delta\text{RI}_{\mathrm{repli}}$ slightly increase.
This confirms that the replication experiment was not successful.

\subsection{WWW-4 Reproducibility Results}

This section corrects the reproducibility results reported in the WWW-4 overview paper~\cite{sakai22www4}.
The target of reproducibility is the run KASYS-CO-REV-6 and the submitted reproducibility run is SLWWW-CO-REP-1.
For reproducibility, we re-evaluate both the original and reproduced runs with the corrected qrels from WWW-4. 
We do not compute any replicability measure because the systems were not tested on a different set of topics as in WWW-3. 

As for WWW-3, reproducibility measures which consider the ordering of documents are not affected by the bug (Figures 2a and 2b in~\cite{sakai22www4}).

\begin{table}[tb]
         \centering
        \caption{Reproducibility results with effectiveness measures: RMSE and p-values. This table replaces Figure~2c of~\citet{sakai22www4}.}
        \begin{tabular}{@{}lll@{}}
            \toprule
                 & RMSE     & $p$-values \\ \midrule
            nDCG & $0.0227$ & $0.5084$  \\
            Q    & $0.0257$ & $0.5856$  \\
            nERR & $0.0595$ & $0.3966$  \\
            iRBU & $0.0123$ & $0.7953$  \\ \bottomrule
        \end{tabular}
        \label{tab:effectiveness_www4}
\end{table}

Table~\ref{tab:effectiveness_www4} reports RMSE and $p$-values for the reproduced run.
The corrected RMSE scores are not very far from those reported in~\cite{sakai22www4} and still support the observation that WWW-4 reproducibility results are better than those reported in other reproducibility experiments~\cite{breuer20,sakai20www3}.
The same applies to $p$-values, even if they tend to be overall lower.

\clearpage

\section{Conclusions}\label{s:conclusions}

This paper corrected the results reported in the NTCIR-14 WWW-2, NTCIR-15 WWW-3, 
and NTCIR-16 overview papers~\cite{mao19,sakai20www3,sakai22www4}.

Our new conclusions are as follows.
\begin{itemize}
\item According to the corrected WWW-2 English run results,
THUIR-E-CO-MAN-Base-3 is statistically the best run in terms of both nDCG and Q.
The same run is also ranked first on average with nERR and iRBU.
(According to the original results,
the same run was statistically the best run in terms of nDCG,
but another THUIR run was statistically the best run in terms of Q.) 
Hence the corrected WWW-2 results make THUIR-E-CO-MAN-Base-3 the clear winner.
(The conclusion in the WWW-2 overview paper was less specific: it said that
``runs from THUIR are the most effective.'')
\item According to the corrected WWW-3 English run results,
mpii-E-CO-NEW-1 and KASYS-E-CO-NEW-\{1,4\}
are statistically the best run in terms of nDCG; the results with Q are similar.
(These three runs also formed the
top cluster according to the original significance test results, 
although KASYS-E-CO-NEW-1
was originally ranked above mpii-E-CO-NEW-1 according to mean scores.)
According to nERR, mpii-E-CO-NEW-1
is statistically the best run, which suggests that
this run is good at navigational searches.
(The same remark was made in the WWW-3 overview paper.)
\item According to the WWW-4 results based on the corrected Gold qrels file,
THUIR-CO-NEW-2 is statistically the best run in terms of nDCG,
but even this run is statistically indistinguishable from
\textbf{KASYS-CO-REV-6}, which is equivalent to 
KASYS-E-CO-NEW-1 from WWW-3.
Hence, we cannot conclude that the improvement over the WWW-3 is substantial.
The results with Q are similar.
(The conclusion in the WWW-4 overview paper remains unchanged.)
In terms of nERR, ORG-TOPICDEV, 
which contains a few relevant documents per topic
found by the Gold assessors at topic development time,
is statistically the best run.
(In the original results, ORG-TOPICDEV did not perform well even in terms of nERR
probably
because the seed relevant documents were not correctly labelled in the Gold qrels file.)
SLWWW-CO-REP-1 performs very similarly
to \textbf{KASYS-CO-REV-6}
and it looks quite successful as a reproducibility run.
(The conclusion in the WWW-4 overview paper remains unchanged.)
\item Reproducibility experiments are affected to a very limited extent. 
The main conclusions from WWW-3 (reproducibility and replicability runs were not successful) and WWW-4 (the reproducibility run was successful to a good extent) still hold.
The only notable difference was observed for the $p$-values of replicability runs in WWW-3, however the corrected $p$-values align better with the $p$-values computed for reproducibility runs. 
\end{itemize}

As a summary of the impact of the bug, 
Table~\ref{t:impact}
shows the Kendall's $\tau$
between the ranking based on the original qrels file and that based on the corrected qrels file for each evaluation and for each task,
with 95\%CIs.
It can be observed that, while the bug did not affect the main conclusions 
in the overview papers substantially,
the impact on the entire system rankings is indeed substantial.
Recall that the low correlation for nERR for the WWW-4 ranking
is due to the fact that 
the seed relevant documents identified by the Gold assessors at topic development time
for the RND-based topics (i.e., one-half of the topic set)
were not recognised as relevant in the original qrels file.

We plan to report on detailed Gold-Bronze and PRI-RND comparisons based on the WWW-4 data elsewhere.

\begin{table}[t]
\begin{center}

\caption{Summary of the impact of the bug: Kendall's $\tau$ (with 95\%CIs)
between the ranking based on the original qrels file and that based on the corrected qrels file for each evaluation and for each task.
}\label{t:impact}

\begin{tabular}{c|c}
\hline
\multicolumn{2}{c}{(a) WWW-2: 80 topics; $n=20$ runs}\\
\hline
nDCG	&0.758 [0.583, 0.866]\\
Q		&0.679 [0.465, 0.818]\\
nERR	&0.653 [0.427, 0.802]\\
iRBU	&0.711 [0.512, 0.838]\\
\hline
\multicolumn{2}{c}{(b) WWW-3: 80 topics; $n=37$ runs}\\
\hline
nDCG	&0.877 [0.813, 0.920]\\
Q		&0.844 [0.766, 0.898]\\
nERR	&0.791 [0.690, 0.862]\\
iRBU	&0.616 [0.457, 0.737]\\
\hline
\multicolumn{2}{c}{(c) WWW-4: 50 topics; $n=18$ runs (Gold assessments)}\\
\hline
nDCG	&0.686 [0.457, 0.830]\\
Q		&0.863 [0.744, 0.929]\\
nERR	&0.340 [0.008, 0.605]\\
iRBU	&0.601 [0.335, 0.778]\\
\hline
\end{tabular}

\end{center}
\end{table}

\section*{Acknowledgements}

Apologies yet again to the WWW participants and the NTCIR chairs for the inconvenience and and confusion caused by the bug.

This paper was partially supported by the the European Union's Horizon 2020 research and innovation programme under the Marie Sk\l{}odowska-Curie grant agreement No. 893667.

\section*{Disclaimer}

Certain companies and products are identified in this paper in order to specify the experimental procedure adequately.  
Such identification is not intended to imply recommendation or endorsement by NIST, nor is it intended to imply that the products or companies identified are necessarily the best available for the purpose.

\appendix

\section*{Appendix: The Effect of Noisy Parts of the Qrels Files on the Official Results}

This appendix discusses the effect of noisy parts the qrels files (due to the aforementioned bug) on the official system rankings,
by comparing the following three types of system ranking in terms of Kendall's $\tau$ on a common topic set.
\begin{description}
\item[good$+$noise] The original system ranking where the qrels file suffers from the bug;
\item[good$+$corrected] The corrected system ranking where the noisy parts of the qrels file were corrected, as we have reported in the main part of this paper;
\item[good$+$NULL] An alternative system ranking where the same noisy parts of the qrels file were \emph{removed} rather than corrected.
\end{description}
For all qrels files, we use the linear gain value setting.
If good$+$noise resembles good$+$corrected more than good$+$NULL does,
this suggests that the noisy parts are somewhat useful for estimating the good$+$corrected ranking.

The good$+$NULL option does not apply to the WWW-4 Gold data,
because each topic has exactly one gold assessor by definition:
each topic is either PRI-based or RND-based,
and removing the noisy RND-based assessments simply
means dropping 25 of the 50 topics.
Hence  the following subsections discuss the WWW-2 and WWW-3 system rankings.


As we have described in Section~\ref{s:www2},
the 80 WWW-2 topics is composed of
27 PRI-PRI topics,
26 RND-RND topics,
and 
27 PRI-RND topics,
and the official PRI-based assessments were the noise for this particular test collection.
Hence, for WWW-2, the effect of noise can  be investigated by focussing on the 27 PRI-RND topics (for which
the original qrels file contains 50\% noise per document),
and comparing the following three types system ranking.
\begin{description}
\item[good$+$noise] A ranking based on the original qrels file with relevance levels L0-L4, 
where the noisy PRI-based assessments are combined with the RND-based assessments;
\item[good$+$corrected] A ranking based on the corrected qrels file with relevance levels L0-L4,
where the above noisy PRI-based assessments have been corrected;
\item[good$+$NULL] A ranking based on a 3-point scale (L0-L2) qrels file that relies only on the RND-based assessments.
\end{description}

As we have described in Section~\ref{s:www3},
each of
the 80 WWW-3 topics rely on 4 PRI-based and 4 RND-based assessments per document
in the original qrels file,
and the RND-based assessments were the noise (i.e., 50\% noise per document for every topic).
Hence, for WWW-3, the effect of noise can be investigated on the full topic set,
by comparing the following three types of system ranking
\begin{description}
\item[good$+$noise] A ranking based on the original qrels file with relevance levels L0-L4;
\item[good$+$noise] A ranking based on the corrected qrels file with relevance levels L0-L4, 
with all four RND-based assessments corrected; and
\item[good$+$noise] A ranking based on a 4-point scale (L0-L3) qrels file that relies only on the 4 PRI-based assessments,
formed by appling the log-based merging scheme described in Section~\ref{s:www3}. 
\end{description}

\begin{table}[t]
\begin{center}

\caption{Effect of noise in the qrels file on the system ranking in terms of Kendall's $\tau$ with 95\%CIs.
}\label{t:noise}

\begin{tabular}{c|c|c}
\hline
\multicolumn{3}{c}{(a) WWW-2 ($n=20$ runs; 27 PRI-RND topics)}\\
\hline
			&good$+$corrected	&good$+$NULL\\
\hline
\multicolumn{3}{c}{nDCG}\\
\hline
good$+$noise		&0.747 [0.566, 0.859]		&0.895 [0.808, 0.944]\\
good$+$corrected	& -			&0.779 [0.616, 0.878]\\
\hline
\multicolumn{3}{c}{Q}\\
\hline
good$+$noise		&0.642 [0.412, 0.795]		&0.779 [0.616, 0.878]\\
good$+$corrected	& -			&0.768 [0.599, 0.872]\\
\hline
\multicolumn{3}{c}{nERR}\\
\hline
good$+$noise		&0.589 [0.338, 0.762]		&0.705 [0.503, 0.834]\\
good$+$corrected	& -			&0.663 [0.442, 0.808]\\
\hline
\multicolumn{3}{c}{iRBU}\\
\hline
good$+$noise		&0.679 [0.465, 0.818]		&0.721 [0.527, 0.844]\\
good$+$corrected	& -			&0.521 [0.248, 0.717]\\
\hline
\multicolumn{3}{c}{(b) WWW-3 ($n=37$ runs; 80 topics)}\\
\hline
			&good$+$corrected	&good$+$NULL\\
\hline
\multicolumn{3}{c}{nDCG}\\
\hline
good$+$noise		&0.877 [0.813, 0.920]		&0.908 [0.859, 0.940]\\
good$+$corrected	& -			&0.947 [0.918, 0.966]\\
\hline
\multicolumn{3}{c}{Q}\\
\hline
good$+$noise		&0.844 [0.766, 0.898]		&0.913 [0.867, 0.944]\\
good$+$corrected	& -			&0.907 [0.858, 0.940]\\
\hline
\multicolumn{3}{c}{nERR}\\
\hline
good$+$noise		&0.791 [0.690, 0.862]		&0.857 [0.784, 0.906]\\
good$+$corrected	& -			&0.928 [0.889, 0.954]\\
\hline
\multicolumn{3}{c}{iRBU}\\
\hline
good$+$noise		&0.616 [0.457, 0.737]		&0.800 [0.703, 0.868]\\
good$+$corrected	& -			&0.779 [0.674, 0.853]\\
\hline
\end{tabular}

\end{center}
\end{table}


Table~\ref{t:noise} shows the results of comparing 
good$+$noise, good$+$corrected, and good$+$NULL for 
the WWW-2 and WWW-3 runs.
Note that the WWW-3 results should be considered more reliable
as it utilises more topics, more runs, and
more relevance labels per document compared to WWW-2.
(For WWW-3,
the comparisons between good$+$noise
and good$+$corrected have already been reported 
in Table~\ref{t:impact},
as we use the full WWW-3 topic set in this additional analysis as well.)
It can be observed that, with the exception of iRBU on the WWW-2 data,
the good$+$NULL ranking 
resembles the good$+$corrected ranking 
more than the good$+$noise ranking does on average in each case.
This suggests that the noisy parts
in the original qrels file are generally not useful for 
estimating the good$+$corrected (i.e., true) ranking.



\bibliographystyle{ACM-Reference-Format}

\bibliography{www234corrected}


\end{document}